\documentclass[12pt]{article}
\usepackage{amsmath}
\usepackage{hhline}
\usepackage{amssymb}
\usepackage{times}
\usepackage{graphicx}
\DeclareGraphicsExtensions{.eps}
\usepackage{psfrag}

\newlength{\dinwidth}
\newlength{\dinmargin}
\newcommand{\ARIADNE}{{\sc Ariadne}}
\newcommand{\DISASTER}{{\sc Disaster}}
\newcommand{\DISENT}{{\sc Disent}}
\newcommand{\DJANGO}{{\sc Django}}
\newcommand{\HERACLES}{{\sc Heracles}}
\newcommand{\HERWIG}{{\sc Herwig}}
\newcommand{\JADE}{{\sc Jade}}
\newcommand{\JETSET}{{\sc Jetset}}

\newcommand{\LEPTO}{{\sc Lepto}}
\newcommand{\MEPJET}{{\sc Mepjet}}
\newcommand{\aeff}{\alpha_{\rm eff}}

\newcommand{\an}{\overline{\alpha}_0}
\newcommand{\ao}{\overline{\alpha}_1}
\newcommand{\ap}{\overline{\alpha}_{p-1}}
\newcommand{\as}{\alpha_s}
\newcommand{\asmz}{\alpha_s(M_Z)}

\newcommand{\chin}{\chi^2/{\rm dof}}
\newcommand{\cm}{\,{\rm cm}}
\newcommand{\gev}{\,{\rm GeV}}
\newcommand{\gevq}{\,{\rm GeV}^2}
\newcommand{\grad}{{^\circ}}
\newcommand{\hftwo}{\hspace*{\fill}}

\newcommand{\hch}{{\scriptscriptstyle h \in {\rm CH}}}

\newcommand{\ee}{\mbox{$e^+e^-$~}}
\newcommand{\mean}[1]{\left< #1 \right>}
\newcommand{\fmean}{\mean{F}}
\newcommand{\fpert}{\fmean^{\rm pert}}
\newcommand{\fpow}{\fmean^{\rm pow}}

\newcommand{\mf}{\mu_{\scriptscriptstyle F}}
\newcommand{\mi}{\mu_{\scriptscriptstyle I}}
\newcommand{\mr}{\mu_{\scriptscriptstyle R}}
\newcommand{\anmi}{\overline{\alpha}_0(\mi=2\gev)}
\newcommand{\order}{{\cal O}}
\newcommand{\pb}{\,{\rm pb}}
\newcommand{\rbthm}{\rule[-2ex]{0ex}{5ex}}

\newcommand{\rbtrr}{\rule[-0.8ex]{0ex}{3.2ex}}

\newcommand{\eprint}[2]{{hep-#1/}#2.}
\newcommand{\eprintk}[2]{{hep-#1/}#2,}
\newcommand{\eprints}[2]{{hep-#1/}#2;\\}

\newcommand{\jrnl}[4]{{#1} {\bf #2} (#3) #4.}
\newcommand{\jrnlk}[4]{{#1} {\bf #2} (#3) #4,}
\newcommand{\jrnls}[4]{{#1} {\bf #2} (#3) #4;\\}
\newcommand{\CPC}{{\em Comp.\ Phys.\ Comm.}}
\newcommand{\EPJC}{{\em Eur.\ Phys.\ J.\ }{\bf C}}
\newcommand{\EPJCd}{{\em Eur.\ Phys.\ J.\ direct }{\bf C}}

\newcommand{\JHEP}{\bf JHEP}
\newcommand{\JPG}{{\em J.~Phys.\ }{\bf G}}

\newcommand{\NIMA}{{\em Nucl.~Instr.\ and Meth.\ }{\bf A}}
\newcommand{\NPB}{{\em Nucl.~Phys.\ }{\bf B}}

\newcommand{\PLB}{{\em Phys.~Lett.\ }{\bf B}}

\newcommand{\PRD}{{\em Phys.~Rev.\ }{\bf D}}
\newcommand{\ZPC}{{\em Z.~Phys.\ }{\bf C}}

\begin{document}

\begin{titlepage}
  
\noindent
{\tt DESY 99-193}\hfill{\tt ISSN 0418-9833}\\
December 1999
  
\vspace{2cm}
  
\begin{center}
  \begin{Large}
    
    {\bf Investigation of Power Corrections to\\[.2em]
      Event Shape Variables measured\\[.2em]
      in Deep-Inelastic Scattering}\\
    
    \vspace{2cm}
    
    H1 Collaboration
    
  \end{Large}
\end{center}

\vspace{2cm}

\begin{abstract}
  \noindent
  Deep-inelastic $ep$ scattering data, taken with the H1 detector at
  HERA, are used to study the event shape variables thrust, jet
  broadening, jet mass, $C$~parameter and two kinds of differential
  two-jet rate.  The data cover a large range of the four-momentum
  transfer $Q$, which is considered to be the relevant energy scale,
  between $7\gev$ and $100\gev$.  The $Q$ dependences of the mean
  values are compared with second order calculations of perturbative
  QCD applying power law corrections proportional to $1/Q^p$ to
  account for hadronization effects. The concept of power corrections
  is investigated by fitting simultaneously a non-perturbative
  parameter $\ap$ and the strong coupling constant
  $\as$.
\end{abstract}

\vspace{1.5cm}

\begin{center}
  To be submitted to Eur.~Phys.~J.~C
\end{center}

%\end{titlepage}
\newpage

%
%          COPY THE AUTHOR AND INSTITUTE LISTS 
%       AT THE TIME OF THE T0-TALK INTO YOUR AREA
%
% from /h1/iww/ipublications/h1auts.tex 

\begin{flushleft}
  %   H1AUTS  Author list by names, no. of authors  344
%           status: 30/07/99   12.55.56
 C.~Adloff$^{33}$,                %WUPP-ST                  Adloff             
 V.~Andreev$^{24}$,               %LPI -PD                  Andreev            
 B.~Andrieu$^{27}$,               %ECPL-PD                  Andrieu            
 V.~Arkadov$^{35}$,               %ZEUT-PD    10/96         Arkadov            
 A.~Astvatsatourov$^{35}$,        %ZEUT-ST     2/98         Astvatsatourov     
 I.~Ayyaz$^{28}$,                 %PARI-ST       5/96       Ayyaz              
 A.~Babaev$^{23}$,                %ITEP-PD                  Babaev             
 J.~B\"ahr$^{35}$,                %ZEUT-PD                  Baehr              
 P.~Baranov$^{24}$,               %LPI -PD                  Baranovp           
 E.~Barrelet$^{28}$
% ...

%%% Local Variables: 
%%% mode: latex
%%% TeX-master: "draft"
%%% End: 

\end{flushleft}

% from /h1/iww/ipublications/h1inst.tex

\begin{flushleft} 
  {\it %     H1 Institutes as appearing on publications
% ...
$ ^{23}$ Institute for Theoretical and Experimental Physics,
          Moscow, Russia \\
 $ ^{24}$ Lebedev Physical Institute, Moscow, Russia$^{f,k}$ \\
% ...
$ ^{27}$ LPNHE, \'{E}cole Polytechnique, IN2P3-CNRS, Palaiseau, France \\
 $ ^{28}$ LPNHE, Universit\'{e}s Paris VI and VII, IN2P3-CNRS,
          Paris, France \\
% ...
$ ^{33}$ Fachbereich Physik, Bergische Universit\"at Gesamthochschule
          Wuppertal, Wuppertal, Germany$^a$ \\
% ...
$ ^{35}$ DESY, Zeuthen, Germany$^a$ \\
% ...
$ ^a$ Supported by the Bundesministerium f\"ur Bildung, Wissenschaft,
        Forschung und Technologie, FRG,
        under contract numbers 7AC17P, 7AC47P, 7DO55P, 7HH17I, 7HH27P,
        7HD17P, 7HD27P, 7KI17I, 6MP17I and 7WT87P \\
% ...
$ ^f$ Supported by the Deutsche Forschungsgemeinschaft \\
% ...
$ ^k$ Supported by Russian Foundation for Basic Research 
       grant no. 96-02-00019 \\
% ...
% $ ^{m}$ Foundation for Polish Science fellow \\

%%% Local Variables: 
%%% mode: plain-tex
%%% TeX-master: "draft"
%%% End: 
 }
\end{flushleft}

\end{titlepage}
%\newpage

\section{Introduction}
\label{introduction}

Hadronic final states in deep-inelastic $ep$ scattering (DIS) $ep
\rightarrow eX$ offer an interesting environment to study
non-perturbative hadronization phenomena and predictions of
perturbative QCD over a wide range of momentum transfer $Q$ in a
single experiment.  A major limitation comes from the treatment of the
hadronization of partons, usually modelled by phenomenological event
generators.  Recent theoretical developments suggest that these
non-perturbative hadronization contributions may be described by
$\order(1/Q^p)$ power law corrections~\cite{dwmodel,webber} with
perturbatively calculable coefficients relating their relative
magnitudes.  Fragmentation models are not needed.

First results on an analysis of mean event shape variables as a
function of $Q$ in terms of power corrections and the strong coupling
constant $\asmz$ have been published by the H1
collaboration~\cite{h1eventshapes}.  It could be shown that power
corrections can be successfully applied to the variables thrust and
jet mass, but they failed to describe the observed jet broadening.  In
order to further test the concept of power corrections, the previous
work is considerably extended in the present paper.  The analyzed
integrated luminosity at high
$Q$ is tripled, the data correction methods are refined and
additional event shape variables are investigated.  Theoretical
progress has come from calculations of two-loop effects and the
problem of jet broadening has been revisited.  A comprehensive study
of power corrections to the mean values of the event shape variables
thrust, jet broadening, jet mass, $C$ parameter and differential
two-jet rates will be presented.  The data cover a large kinematical
phase space of $7\gev < Q < 100\gev$ in momentum transfer and $0.05 <
y < 0.8$ in the inelasticity $y$.

%%% Local Variables: 
%%% mode: latex
%%% TeX-master: "draft"
%%% End: 
     % Introduction
\section{Event Shapes}
\label{eventshapes}

\subsection{The Breit Frame}
\label{breitframe}

Event shape analyses in deep-inelastic scattering are based on the
observation of the hard scattering of a quark or gluon which has to be
isolated from the target (proton remnant) fragmentation region.  A
particularly suitable frame of reference to study the current region
with minimal contamination from target fragmentation effects is the
Breit frame.  Consider, for illustration, $ep$ scattering in the quark
parton model.  In the Breit system the purely space-like gauge boson
$\gamma/Z$ with four-momentum $q = \{0,0,0,-Q\}$ collides with the
incoming quark with longitudinal momentum $p^{\rm in}_{q\,z} = Q/2$.
The outgoing quark is back-scattered with longitudinal momentum
$p^{\rm out}_{q\,z} = -Q/2$ while the proton fragments in the opposite
hemisphere carrying longitudinal momentum $p_{r\,z} =
Q/2\cdot(1-x)/x$, where $x$ is the fractional momentum of the struck
quark in the proton with momentum $P$.  Employing the boson direction
as a `natural' axis, the boost into the Breit frame, defined by
$2x{\bf P}+{\bf q}={\bf 0}$, thus provides a clean separation into a
current and a target hemisphere and may be used to classify event
topologies at a scale $Q$.  Higher order processes like QCD Compton
scattering and boson gluon fusion modify this simple picture.
However, without knowledge of the detailed structure of the hadronic
final state, e.g.\ jet directions, the proton remnant is maximally
separated from the current fragmentation.

The kinematic quantities needed to perform the Breit frame
transformation are calculated from the scattered lepton
($E_{e'},\theta_{e'}$) and hadron
measurements ($E_i,\theta_i$) where $i$ runs over all hadronic objects:%
\footnote{Polar angles $\theta$ are defined with respect to the
  incident proton direction.}
\begin{eqnarray}
  Q^2 & = & 4E_eE_{e'}\cos^2\frac{\theta_{e'}}{2}\,,\\[.5em]
  y \equiv y_e & = & 1 - \frac{E_{e'}}{E_e}\sin^2\frac{\theta_{e'}}{2}  
  \quad\quad\quad\quad {\rm for} \quad y_e > 0.15\,,\\[.5em]
  y \equiv y_h & = & \frac{\sum\limits_i E_i(1-\cos\theta_i)}{2E_e} 
  \quad\quad\quad\,\, {\rm for} \quad y_e < 0.15\,,
\end{eqnarray}
with $E_e = 27.5\gev$ and $E_p = 820\gev$ being the beam energies.
The inelasticity $y = y_e$ is chosen for sufficiently large values.
However, since the resolution in $y_e$ degrades severely at low
values, $y = y_h$ is taken if $y_e < 0.15$.  This procedure ensures
least uncertainty in the Lorentz transformation to the Breit frame.

\subsection{Definition of Event Shape Variables}
\label{definitions}

The dimensionless event shape variables thrust, jet broadening, $C$
parameter and jet mass are studied in the current hemisphere (CH) of
the Breit system.  The sums extend over all particles $h$ of the
hadronic final state in the CH with four-momenta $p_h = \{E_h,{\bf p}_h\}$.
The current hemisphere axis ${\bf n} = \{0,0,-1\}$ coincides with the
boson direction.  The following
collinear- and infrared-safe definitions of event shape variables%
\footnote{{\em Note:}\/ The notation of event shape variables is
  different from the previous analysis~\cite{h1eventshapes}, but more
  transparent.  All indices are dropped except for $\tau_C$.  The
  normalization is always performed with respect to the sum of momenta
  or the total energy in the current hemisphere and not $Q$ as was done
  previously for $\rho$.}  are used:

{\em Thrust $\tau \equiv 1 - T$}\/ measures the longitudinal momentum
components projected onto the current hemisphere axis
\begin{eqnarray}
  \tau & = & 1 - \frac{\sum\limits_\hch |{\bf p}_h\cdot {\bf n}|} 
  {\sum\limits_\hch |{\bf p}_h|} 
  \ = \ 1 - \frac{\sum\limits_\hch |{\bf p}_{z\,h}|}
  {\sum\limits_\hch |{\bf p}_h|}\,.
  \label{eqn:thrust} 
\end{eqnarray}

{\em Thrust $\tau_C \equiv 1 - T_C$}\/ uses the direction ${\bf n}_T$
which maximizes the sum of the longitudinal momenta of all particles
in the current hemisphere along this axis
\begin{eqnarray}
  \tau_C & = &
  1-\max_{\scriptscriptstyle{\bf n'},{\bf n'}^2=1}
  \frac{\sum\limits_\hch |{\bf p}_h\cdot {\bf n'}|} 
  {\sum\limits_\hch |{\bf p}_h |} =
  1-\frac{\sum\limits_\hch |{\bf p}_h\cdot {\bf n}_T|} 
  {\sum\limits_\hch |{\bf p}_h |} \,.
  \label{eqn:thrustc}
\end{eqnarray}
This definition is analogous to that used in \ee\ experiments and
represents a mixture of longitudinal and transverse momenta with
respect to the boson axis.

The {\em Jet Broadening $B$}\/ measures the scalar sum of transverse
momenta with respect to the current hemisphere axis
\begin{eqnarray}
  B & = & \frac{\sum\limits_\hch |{\bf p}_h\times {\bf n}|} 
  {2\,\sum\limits_\hch |{\bf p}_h|}
  \ = \ \frac{\sum\limits_\hch |{\bf p}_{\perp\,h}|}
  {2\,\sum\limits_\hch |{\bf p}_h|}\,.
  \label{eqn:bparameter}
\end{eqnarray}

The squared {\em Jet Mass $\rho$}\/ is normalized to four times the squared
total energy in the current hemisphere
\begin{eqnarray}
  \rho & = & \frac{(\sum\limits_\hch p_h )^2}
                  {(2\,\sum\limits_\hch E_h )^2 }\,.
  \label{eqn:jetmass} 
\end{eqnarray}

The {\em $C$ Parameter}\/ is derived from the eigenvalues $\lambda_i$
of the linearized momentum tensor $\Theta^{jk}$
\begin{eqnarray}
  \Theta^{jk} & = & \frac{\sum\limits_\hch p^j_h p^k_h
    / |{\bf p}_h|}{\sum\limits_\hch |{\bf p}_h|}\nonumber\\[.5em]
  C & = & 3(\lambda_1\lambda_2 + \lambda_2\lambda_3 + \lambda_3\lambda_1)\,.
  \label{eqn:cparameter}
\end{eqnarray}
The real symmetric matrix $\Theta^{jk}$ has eigenvalues $\lambda_i$
with $0 < \lambda_3 \leq \lambda_2 \leq \lambda_1 < 1$. It describes
an ellipsoid with orthogonal axes named minor, semi-major and major
corresponding to the three eigenvalues. The major axis is similar but not
identical to ${\bf n}_T$. If all momenta are collinear then two
eigenvalues and hence $C$ are equal to zero.

Higher order processes may lead to event configurations where the
partons are scattered into the target hemisphere and the current
hemisphere may be completely empty except for migrations due to
hadronization fragments.  In order to be insensitive to such effects
and to keep the event shape variables infrared safe~\cite{seymour} the
total available energy in the current hemisphere has to exceed $20\%$
of the value expected in the quark parton model
\begin{eqnarray}
  E_{CH} \equiv \sum\limits_\hch E_h & > & Q/10\,.
  \label{eqn:echcut} 
\end{eqnarray}
Otherwise the event is ignored.  This cut-off is part of the event
shape definitions, its precise value is not critical.

The event shapes defined in the current hemisphere may be
distinguished according to the event axis used.  Thrust $\tau$ and the
jet broadening $B$ employ momentum vectors projected onto the boson
direction. Thrust $\tau_C$ and the $C$ parameter calculate their own
axis, while the jet mass $\rho$ does not depend on any event
orientation.

Another class of event shapes investigates the number of $(n + 1)$
jets found in an event, where $+1$ denotes the proton remnant.  Jets
are searched for in the complete accessible phase space, i.e.\ in both
the current and target hemispheres of the Breit frame.  Two schemes of
jet definitions are applied: the Durham or $k_t$
algorithm~\cite{ktalgorithm} and a factorizable \JADE\ 
algorithm~\cite{jadealgorithm} adapted to DIS\@.  Both jet finding
procedures introduce two distance measures: one for distances between
two four-vectors, $y_{ij}$, and another one for the separation of each
particle from the remnant, $y_{ir}$.  The following distance measures
are used:
\begin{eqnarray}
  \mbox{Durham or $k_t$ algorithm}\quad\quad\quad\quad\quad & & \nonumber\\
  y_{ij} & = & \frac{2\min(E_i^2,E_j^2)(1-\cos\theta_{ij})}{Q^2}\,,\\[.5em]
  y_{ir} & = & \frac{2E_i^2(1-\cos\theta_{ir})}{Q^2}\,,\\[1ex]
  \mbox{factorizable \JADE\ algorithm}\quad\quad\quad & & \nonumber\\
  y_{ij} & = & \frac{2E_iE_j(1-\cos\theta_{ij})}{Q^2}\,,\\[.5em]
  y_{ir} & = & \frac{2E_ixE_p(1-\cos\theta_{ir})}{Q^2}\,,
\end{eqnarray}
where $\theta_{kl}$ is the angle between the two momentum vectors.
Since the direction of the proton remnant coincides with the $+z$ axis
for the coordinate system chosen here $\theta_{ir}$ simplifies to the
polar angle $\theta_{i}$.  The pair with the minimal $y_{ij}$ or $y_{ir}$
value of all
possible combinations is selected to either form a new pseudo-particle
vector or to assign the particle $i$ to the remnant.  The whole
procedure is repeated until a certain number of jets is found.  The
event shape variables $y_{k_t}$ ($k_t$ algorithm) and $y_{fJ}$
(factorizable \JADE\ algorithm) are defined as that $y$ value $y_{ij}$
or $y_{ir}$ where the transition from $(2+1) \rightarrow (1+1)$ jets
occurs.

Throughout the paper the symbol $F$ will be used as a generic name for
any event shape variable defined above. Note that for all of them
$F\rightarrow 0$ in case of quark parton model like reactions.
Theoretical calculations of event shape distributions and means will
be discussed in section~\ref{qcdframework}.

%%% Local Variables: 
%%% mode: latex
%%% TeX-master: "draft"
%%% End: 
      % Definition of the Event Shapes
\section{H1 Detector and Event Selection}

\subsection{The H1 Detector}
\label{h1det}

Deep-inelastic $ep$ scattering events were collected during the years
$1994-1997$ with the H1 detector~\cite{h1det} at HERA\@.  Electrons or
positrons with $E_e = 27.5\gev$ collide with $E_p = 820\gev$ protons
at a center of mass energy of $\sqrt{s} = 300\gev$.  Only calorimetric
information is used to reconstruct the final state.  The direction of
the scattered lepton and the event vertex are obtained by exploiting
additional information from the tracking detectors.  The calorimeters
cover the polar angles $4\grad \le \theta \le 176\grad$ and the full
azimuth.

The calorimeter system consists of a liquid argon (LAr) calorimeter, a
backward calorimeter and a tail catcher (instrumented iron yoke).  The
LAr sampling calorimeter ($4\grad \le \theta \le 154\grad$) consists
of a lead/argon electromagnetic section and a stainless steel/argon
section for the measurement of hadronic energy.
A detailed {\em in situ}\/ calibration provides the accurate energy
scales.  The lepton energy uncertainty in the LAr calorimeter varies
between $1\%$ in the backward region and $3\%$ in the forward region.
The systematic uncertainty of the hadronic energy amounts to $4\%$.  A
lead/scintillator electromagnetic backward calorimeter (BEMC) extends
the coverage at large angles ($155\grad \le \theta \le 176\grad$) and
is used to measure the lepton at $Q \le 10 \gev$ with a precision of
$1\%$ for the absolute calibration.
Since $1995$ the backward region has been equipped with a
lead/scintillating fibre calorimeter improving the uncertainty in the
measurement of hadronic energies in the backward region from $15\%$ to
$7\%$. The instrumented iron flux return yoke is used to measure the
leakage of hadronic showers.

Located inside the calorimeters is a tracking system which consists of
central drift and proportional chambers ($25\grad \le \theta \le
155\grad$), a forward track detector ($7\grad \le \theta \le 25\grad$)
and a backward proportional chamber ($155\grad \le \theta \le
175\grad$).  In $1995$ the latter was replaced by backward drift
chambers.  The direction of the scattered lepton is determined by
associating tracking information with the corresponding
electromagnetic cluster.  The lepton scattering angle is known to
within $3\,{\rm mrad}$.  The tracking chambers and calorimeters are
surrounded by a superconducting solenoid providing a uniform field of
$1.15\,{\rm T}$ inside the tracking volume.

\subsection{Event Selection}
\label{eventselection}

The DIS data are divided into a low $Q$ event sample ($Q = 7-10\gev$,
lepton detected in BEMC) and a high $Q$ event sample ($Q =
14-100\gev$, lepton detected in LAr calorimeter) which in turn are
subdivided further into eight bins in $Q$: $7-8\gev$, $8-10\gev$,
$14-16\gev$, $16-20\gev$, $20-30\gev$, $30-50\gev$, $50-70\gev$ and
$70-100\gev$. The following event selection criteria ensure a good
measurement of the final state and a clean data sample:
\begin{enumerate}
\item The energy of the isolated scattered lepton has to exceed
  $E_{e'} > 14\gev$ within $157\grad < \theta_{e'} < 173\grad$ for the
  low $Q$ sample and $E_{e'} > 11\gev$ within $30\grad < \theta_{e'} <
  150\grad$ for the high $Q$ sample respectively.  The calorimetric
  lepton trigger efficiencies are above $99\%$~\cite{h1det,h1highq2}.
\item The inelasticity $y$ is well measured by requiring $0.05
  < y_e < 0.8$ (using the lepton) and $0.05 < y_h$ (using the hadronic
  energy flow).  This criterion suppresses photoproduction events with
  a misidentified lepton.
\item The `quark' direction as calculated from the scattered lepton in
  the quark parton model corresponds to the $-z$ axis of the Breit
  frame.  A minimal value of $\theta_q > 20\grad$ in the laboratory
  system ensures a sufficient detector resolution in polar angle after
  transformation into the Breit frame.
\item A minimal energy in the Breit current hemisphere of $Q/10$ (see
  eq.~(\ref{eqn:echcut}) of section~\ref{definitions}) is essential to
  keep the event shapes $\tau$, $B$, $\tau_C$, $\rho$ and $C$ infrared
  safe. This is part of their {\em definition}.
\item To avoid unphysical peaks at zero for $\tau_C$, $\rho$ and $C$
  at least two hadronic objects are required.  Events containing only
  one such object are {\em not}\/ quark parton model like but are due
  to leakage into the current hemisphere.
\item The total energy in the forward region $(4\grad <\theta <
  15\grad)$ has to be larger than $0.5\gev$ to reduce the proportion
  of diffractive events which are not included into the theoretical
  description of the data.
\item Hadron clusters have to be contained in the calorimeter
  acceptance of $5.7\grad < \theta_{h} < 170\grad$ avoiding the edges
  close to the beam pipe.  The hadronic energy measured in the
  backward region $\theta_h \geq 170\grad$ has to be less than
  $10\gev$ in order to exclude poor measurements.
\item The total longitudinal energy balance must satisfy $30\gev <
  \sum_iE_i(1 - \cos\theta_i) < 65\gev$ in order to suppress initial
  state photon radiation and to further reduce photoproduction
  background.
\item The total transverse momentum has to be $|\vec{p}_\perp| <
  7.5\gev$ (low $Q$ sample) and $|\vec{p}_\perp| < 15\gev$ (high $Q$
  sample), respectively, in order to exclude badly measured events.
\item The energy measurement of the lepton has to be consistent with
  that derived from the double angle method~\cite{dakine} $|(E_{e'} -
  E_{da}) / E_{da}| < 0.25$ in order to further suppress events
  strongly affected by QED radiation.  $E_{da}$ is calculated from the
  directions of the lepton and the hadronic energy flow.
\item An event vertex has to exist within $3\sigma$ of the nominal $z$
  position of the interaction point $|z_v - \langle z_v \rangle| <
  35\cm$.
\item Leptons pointing to dead regions of the LAr calorimeter, i.e.
  $\pm2\grad$ around $\phi$-cracks between modules or $\pm5\cm$ around
  $z$-cracks between wheels, are rejected in order to ensure a
  reliable measurement.
\end{enumerate}

\begin{figure}[htb] 
  \centering \includegraphics{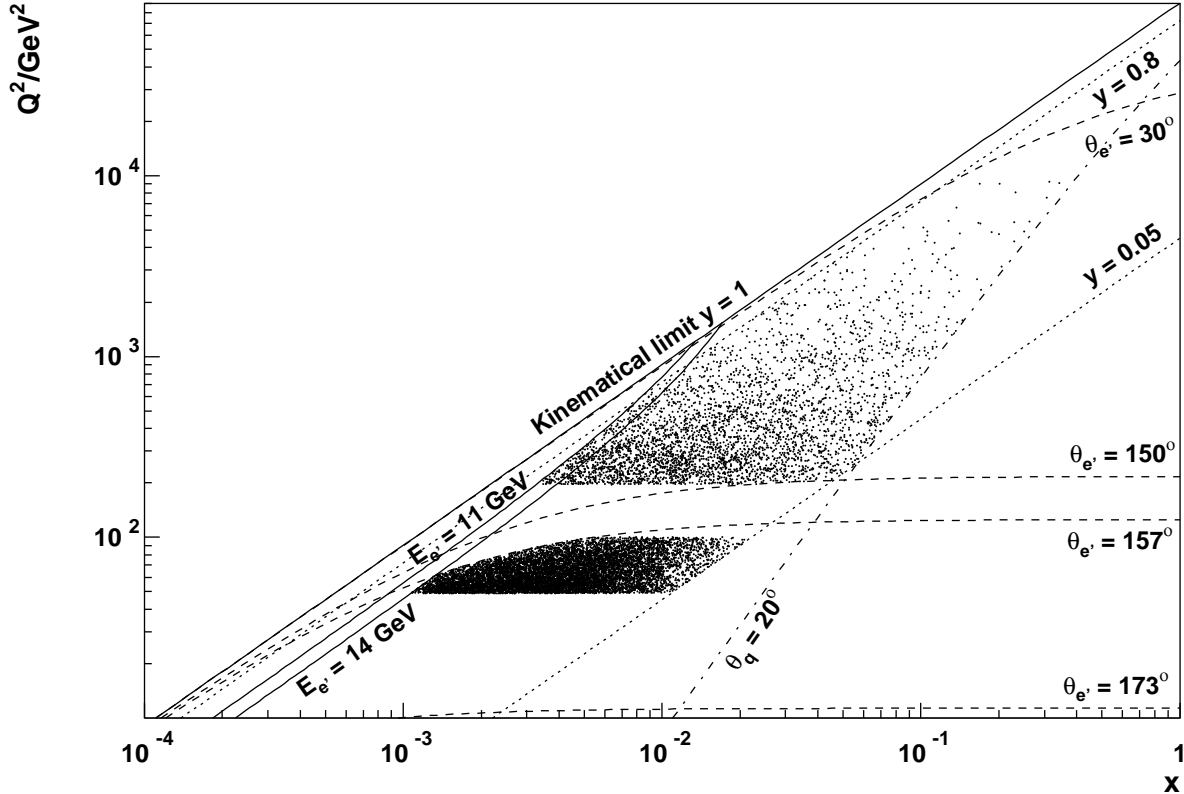}
  \caption{Distribution of selected events in the $x - Q^2$ plane.
    Only a small part of the data corresponding to a luminosity of
    $3.2\pb^{-1}$ is shown. The curves indicate the phase space cuts
    in $E_{e'}$, $\theta_{e'}$, $y$ and $\theta_q$.}
  \label{fig:kinescat}
\end{figure}

The event selection criteria can be separated into phase space cuts,
nos.~$1-4$, representing the common requirements for data and theory,
and data quality cuts, nos.~$5-12$.  Note that the cuts
nos.~$5$ and~$6$ are always applied except for the perturbative
QCD calculations described in section~\ref{qcdframework} where they do
not make sense.  Depending on the theoretical model to compare with they
may be considered as phase space cuts as well.  Not all cuts affect both
the low and high $Q$ data samples, as can be seen from the distribution of
events in the $x - Q^2$ plane shown in figure~\ref{fig:kinescat}.

The contamination from photoproduction background is estimated to be
less than $3\%$ in the low $Q$ sample and negligible at higher values
of $Q$.  Residual radiative effects are accounted for by the data
correction procedure described in section~\ref{measurement}.

The final data samples consist of $9,761$ events at $Q = 7-10\gev$
taken in $1994$ with an integrated luminosity of ${\cal L} =
3.2~\pb^{-1}$ and $42,607$ events at $Q = 14-100\gev$ corresponding to
${\cal L} = 38.2~\pb^{-1}$ recorded from $1994-1997$.

%%% Local Variables: 
%%% mode: latex
%%% TeX-master: "draft"
%%% End: 
        % H1 Detector and Event Selection
\section{Measurement of Event Shapes}
\label{measurement}

The aim of the analysis is to present measured event shape variables
and to compare with second order calculations of perturbative QCD
(pQCD) supplemented with analytical power-law corrections to account
for hadronization effects.  The experimental observables need to be
corrected for detector effects and QED radiation.  Such a procedure
involves Monte Carlo event generators and hence introduces some model
dependencies which are taken into account in the analysis of
systematic uncertainties. They are much smaller than those obtained in
approaches which try to unfold from the data to a `partonic final state'
which is directly comparable to pQCD\@.

\subsection{Event Simulation}

A detailed detector simulation is done by using event samples
generated with the \DJANGO~\cite{django} Monte Carlo program.  The
modelling of parton showers involves a colour dipole model as
implemented in \ARIADNE~\cite{ariadne}.  Alternatively, \DJANGO\ in
combination with \LEPTO~\cite{lepto} without soft colour interactions
has been used to investigate the model dependence.%
\footnote{Including soft colour interactions in the \LEPTO\ 
  version employed spoils the description of the H1 data.}  The hadronization
of partons uses the \JETSET~\cite{jetset} string fragmentation.  QED
radiation on the lepton side, including real photons and virtual
1-loop corrections, is treated by \HERACLES~\cite{heracles}.  Both DIS
packages provide a reasonable description of the measured event shape
distributions and are used to correct the data.  The event generators
\LEPTO\ with soft colour interactions and \HERWIG~\cite{herwig} as
stand-alone programs without QED radiation differ considerably in
their predictions for the hadronic final state and serve to
cross-check the unfolding methods.

\subsection{Data Correction Procedure}

The data analysis proceeds in two steps.  First the data are corrected
for detector effects within the phase space described in
section~\ref{eventselection} applying different techniques.  In a
second step QED radiation and acceptance corrections due to the beam
hole (cut no.~7) are taken into account.

The most reliable method considered to correct event shape
distributions for detector effects is found to be a Bayesian unfolding
procedure~\cite{bayes}.  This technique exploits Bayes' theorem on
conditional probabilities to extract information on the underlying
distribution from the observed distribution.  Although some {\em a
  priori}\/ knowledge on the initial distribution is required ---
it may even be assumed to be uniform in the case of complete ignorance ---
the iterative procedure is very robust and converges to stable results
within three steps.  The program takes correlations properly into
account.

Alternatively, the matrix method employed in the previous
publication~\cite{h1eventshapes} and simple correction factors applied
either bin-by-bin to the distributions or directly to the mean values
have been used. They serve to estimate systematic effects.  The
performance of these correction techniques is checked by studying the
spectra and mean values of the event shapes when unfolding one Monte
Carlo with another and vice versa for various combinations of event
generators. In general, the Bayes method gives the best results; for
details see~\cite{rabbertz}.

The remaining corrections account for QED radiation effects and beam
hole losses.  They are applied on a bin-by-bin basis.  Non-radiative
events are generated using \DJANGO\ with exactly the same conditions
as before except for the radiative effects being switched off.  A
detector simulation is not required. The corrections are based on the
predictions for the hadronic final state within the kinematic phase
space.

The derived bin-to-bin correction factors of the event shape spectra
are close to one except for $\tau$, $B$ and $y_{k_t}$.  For $\tau$ and
$B$, which are defined with respect to the boson axis, radiative
effects are important and non-negligible.  The differential two-jet
rate $y_{k_t}$ is the only variable which is sensitive to acceptance
losses, particularly at low $Q$.  All other event shapes are almost
unaffected by the beam hole cut as expected for the variables defined
in the current hemisphere.  Both Monte Carlo samples, i.e.\ 
\DJANGO/\ARIADNE\ and \DJANGO/\LEPTO, give consistent results.

\subsection{Results on Event Shape Measurements}
\label{finalmeans}

The data correction is performed with the \ARIADNE\ event generator as
implemented in the \DJANGO\ Monte Carlo program.  The final results
are based on the differential distributions obtained with the Bayes
unfolding and subsequent bin-to-bin radiative correction from which
the mean values are calculated.  The total experimental uncertainties
of both the spectra and mean values are evaluated in the same way.
The statistical uncertainties include data as well as Monte Carlo
statistics.  The systematic uncertainties are estimated by comparing
the bin contents and mean of the `standard' distribution with the
values obtained under different conditions or assumptions.  The
effects of various correction procedures and the knowledge of the
calorimeter energy scales are considered.  Systematics from the model
dependence of the Monte Carlo simulation are negligible for the event
shapes defined in the current hemisphere alone.  In case of the $y$
variables the results achieved with \DJANGO/\ARIADNE\ and
\DJANGO/\LEPTO\ are averaged and half of the spread is taken as an
estimate of the uncertainty.

Unfolding uncertainties, being in general asymmetric, are estimated to
be half the maximal deviation to larger and smaller values
respectively due to the alternative unfolding procedures. 
The sum of upper and lower deviation corresponds to half the total spread
and so is somewhat smaller than twice the standard deviation of a uniform
distribution.
The influence of the energy scale uncertainties is taken into account
by repeating the whole analysis and scaling both the electromagnetic
and hadronic energies separately upwards and downwards by the
appropriate amount.  The discrepancies with respect to the central
values are attributed to two further asymmetric systematic
uncertainties.  All three (four in case of $y$) error sources, i.e.\ 
the unfolding bias, the two energy scales and the model dependence,
added in quadrature yield the total systematics.

For the event shapes $\tau$, $B$, $\tau_C$, $\rho$ and $C$, the lepton
energy uncertainty, which directly affects the boost into the Breit
frame, is the largest individual contribution, followed by unfolding
effects.  This can be understood because hadronic systematics cancel
between numerator and denominator for these variables.  The situation
is reversed for $y_{fJ}$ and $y_{k_t}$.  Here, no cancellation occurs
and the systematic uncertainty due to hadronic energies is the larger
of the two energy scale uncertainties.

\begin{figure}[p]
  \centering
  \includegraphics{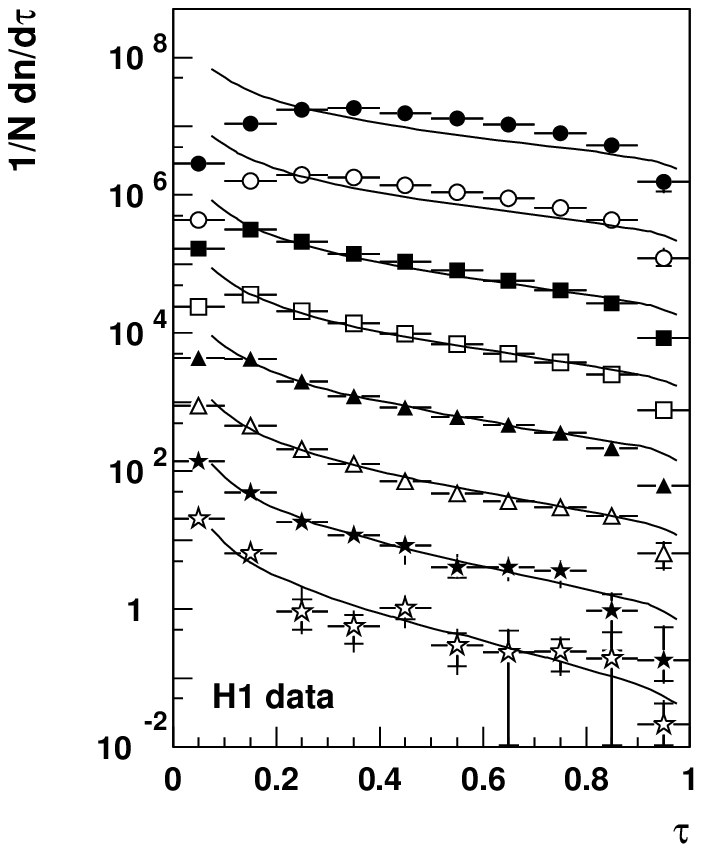}\hftwo%
  \includegraphics{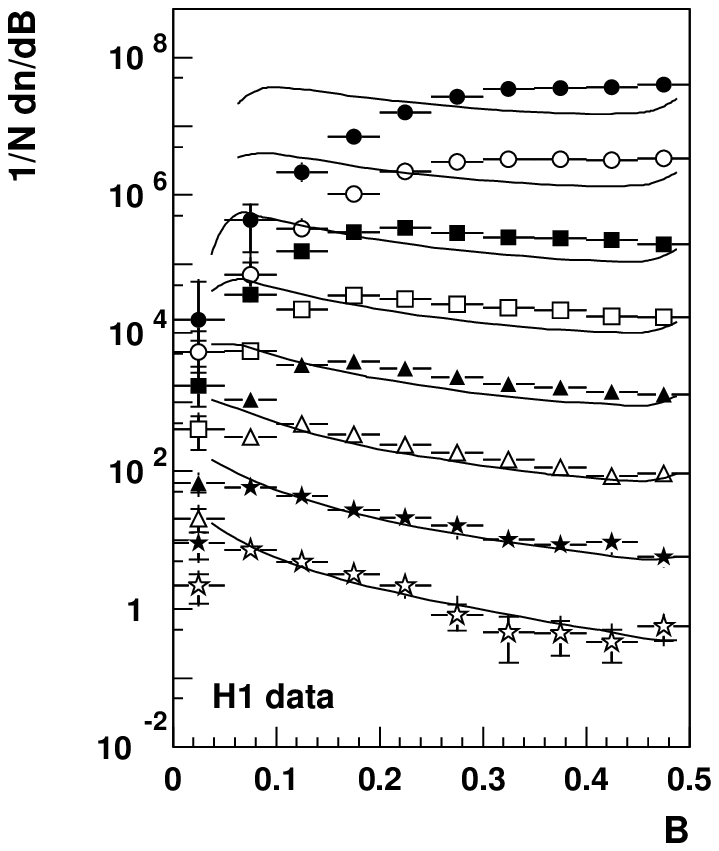}
  \includegraphics{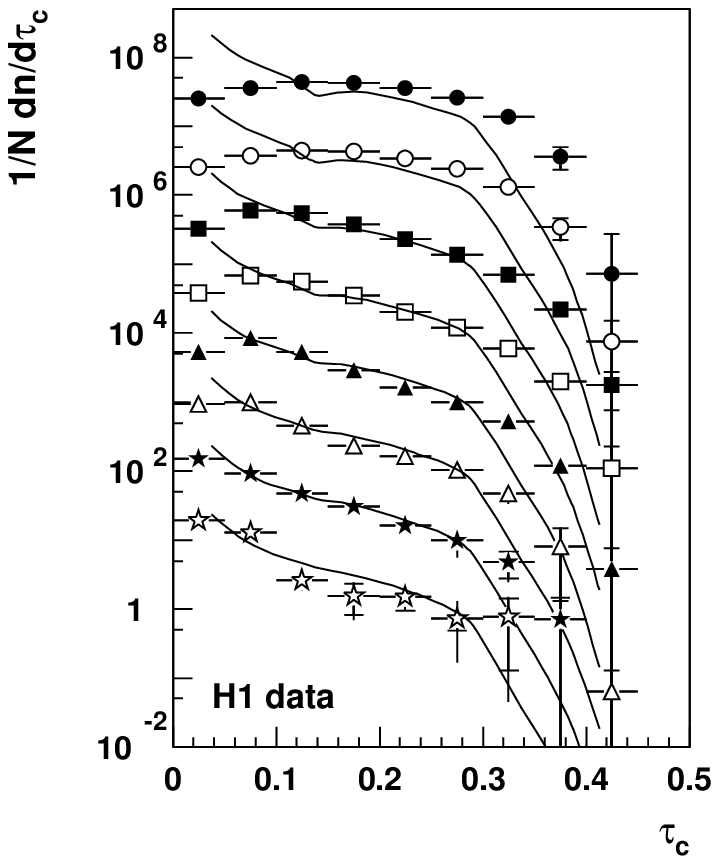}\hftwo%
  \includegraphics{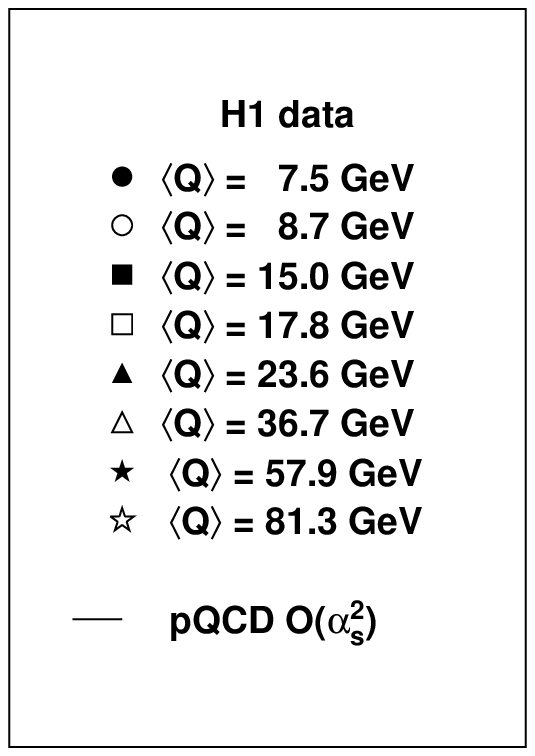}
  \caption{Normalized differential distributions of the event shapes
    $\tau$, $B$ and $\tau_C$.  H1~data (symbols) are compared with
    \DISENT\ NLO calculations (curves) using the MRSA' parton density
    functions with $\asmz = 0.115$.
    The error bars represent statistical and systematic
    uncertainties.  The spectra given at $\mean{Q} = 7.5\gev, \;
    8.7\gev, \ 15.0\gev, \ 17.8\gev, \ 23.6\gev, \ 36.7\gev, \ 
    57.7\gev$ and $81.3\gev$ (from top to bottom) are multiplied by
    factors of $10^n \ (n=7, \ldots, 0)$.}
  \label{fig:dndFhl1}
\end{figure}

\begin{figure}[p] 
  \centering
  \includegraphics{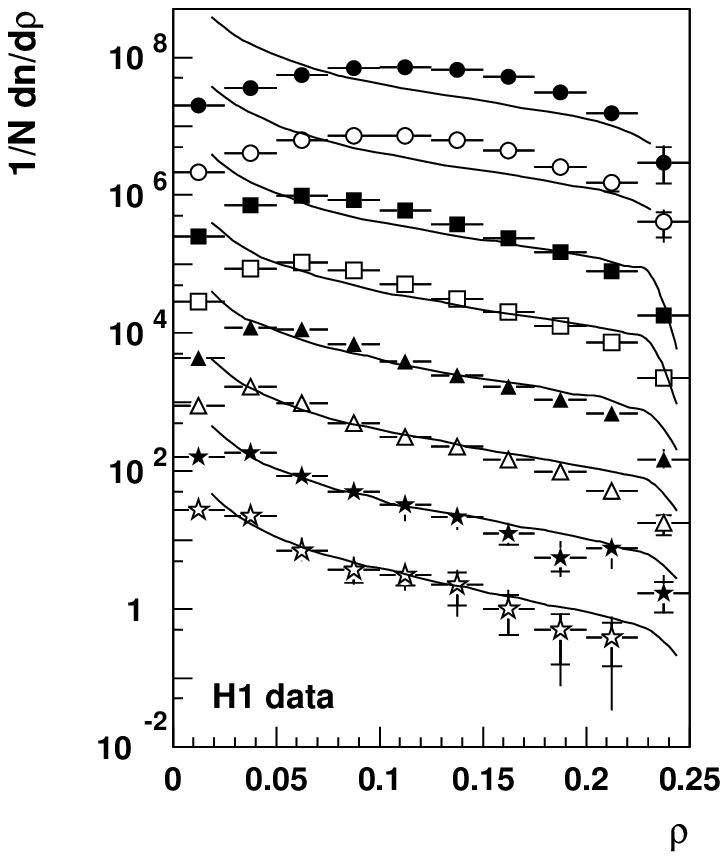}\hftwo%
  \includegraphics{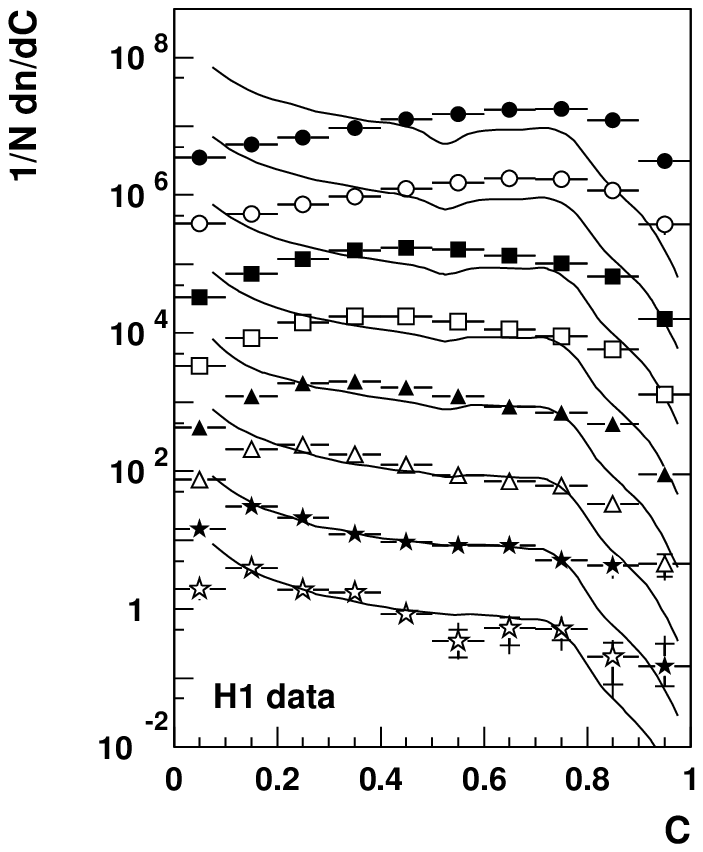}
  \includegraphics{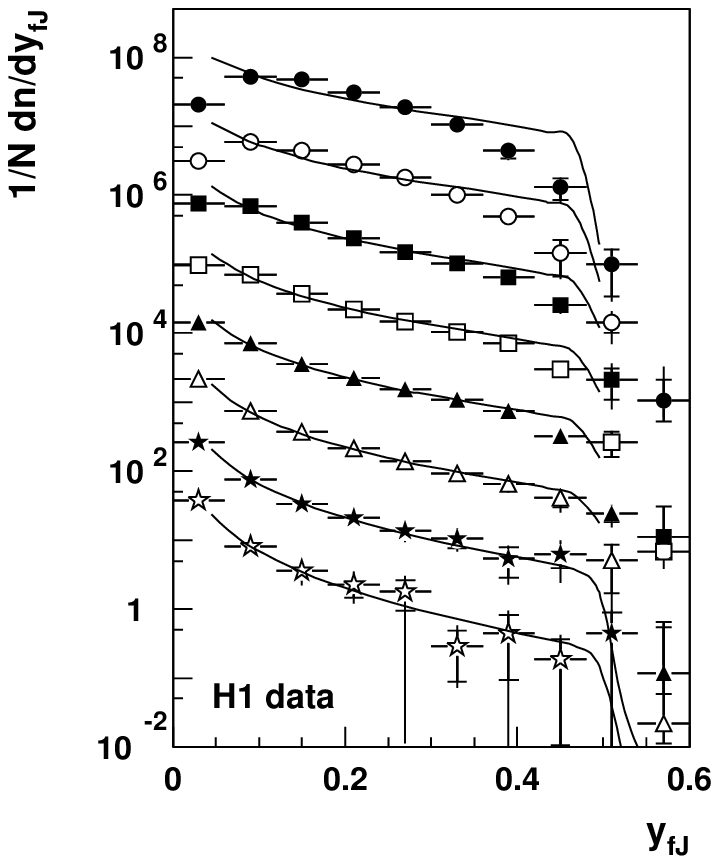}\hftwo%
  \includegraphics{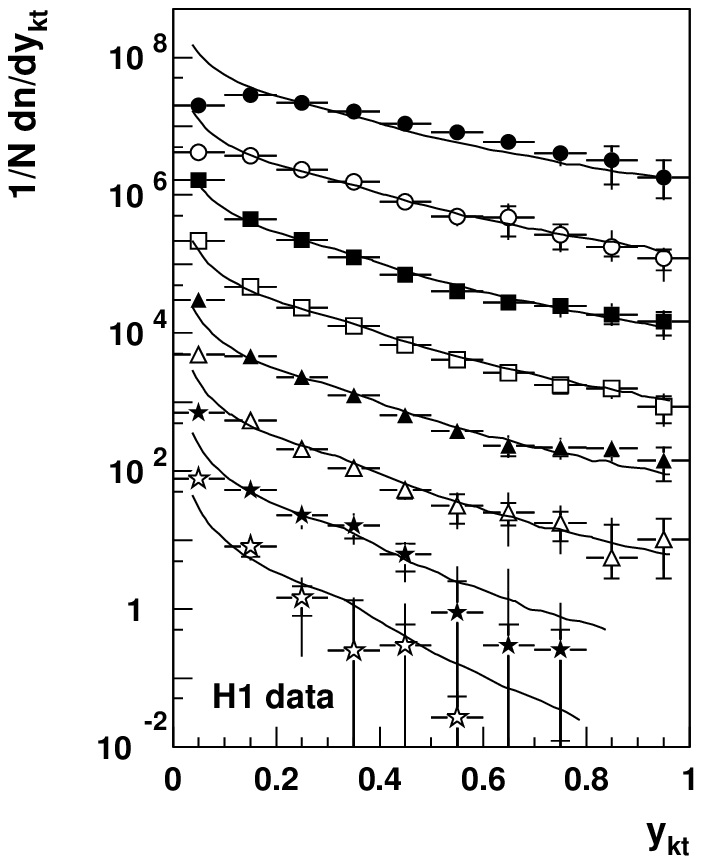}
  \caption{Normalized differential distributions of the event shapes $\rho$,
    $C$, $y_{fJ}$ and $y_{k_t}$.  H1~data (symbols, see
    fig.~\ref{fig:dndFhl1}) are compared with \DISENT\ NLO
    calculations (curves) using the MRSA' parton density functions 
    with $\asmz = 0.115$.  The error
    bars represent statistical and systematic uncertainties. The
    spectra given at $\mean{Q} = 7.5\gev, \; 8.7\gev, \ 15.0\gev, \ 
    17.8\gev, \ 23.6\gev, \ 36.7\gev, \ 57.7\gev$ and $81.3\gev$ (from
    top to bottom) are multiplied by factors of $10^n \ (n=7, \ldots,
    0)$.}
  \label{fig:dndFhl2}
\end{figure}

The corrected event shape distributions are shown in
figures~\ref{fig:dndFhl1} and~\ref{fig:dndFhl2} over a wide range of
$\mean{Q} = 7.5 - 81.3\gev$.  Although the shapes of the spectra for
each variable $F$ are quite different, their common feature is that
they all develop to narrower distributions shifted towards lower
values of $F$ as the available energy $Q$ increases.  It demonstrates
that the events become more collimated.  A comparison with the
predictions of pQCD reveals serious discrepancies especially at low
$Q$ leading to the necessity to include hadronization corrections.
Such corrections are discussed in the next section for the mean values
of the distributions.

The measured mean values are listed in table~\ref{tab:finalmeans}.
Note that the quoted uncertainties do not give the correlations due to
common systematic error sources.  Such correlations are, however,
properly taken into account in the QCD analysis (see
section~\ref{syserrors}).

\begin{table}
  \centering
  \hftwo\begin{tabular}{|c||l|l|}
    \hline
    \multicolumn{1}{|c||}{$\rbthm\mean{Q}/\gev$}&
    \multicolumn{1}{|c|}{$\mean{\tau}$}&
    \multicolumn{1}{|c|}{$\mean{B}$}
    \\\hline\hline
    $7.5$ &
    $0.4402\pm 0.0082~^{+0.0111}_{-0.0122}$&
    $0.3624\pm 0.0034~^{+0.0046}_{-0.0038}$
    \rbtrr\\\hline 
    $8.7$ &
    $0.4017\pm 0.0090~^{+0.0196}_{-0.0080}$&
    $0.3435\pm 0.0040~^{+0.0076}_{-0.0037}$
    \rbtrr\\\hline
    $15.0$&
    $0.3052\pm 0.0034~^{+0.0081}_{-0.0075}$&
    $0.2921\pm 0.0017~^{+0.0044}_{-0.0033}$
    \rbtrr\\\hline
    $17.8$&
    $0.2762\pm 0.0029~^{+0.0091}_{-0.0059}$&
    $0.2760\pm 0.0016~^{+0.0054}_{-0.0034}$
    \rbtrr\\\hline
    $23.6$&
    $0.2279\pm 0.0031~^{+0.0125}_{-0.0071}$&
    $0.2452\pm 0.0018~^{+0.0078}_{-0.0042}$
    \rbtrr\\\hline
    $36.7$&
    $0.1814\pm 0.0049~^{+0.0107}_{-0.0065}$&
    $0.2094\pm 0.0031~^{+0.0083}_{-0.0053}$
    \rbtrr\\\hline
    $57.9$&
    $0.1330\pm 0.0089~^{+0.0092}_{-0.0092}$&
    $0.1717\pm 0.0062~^{+0.0109}_{-0.0096}$
    \rbtrr\\\hline
    $81.3$
    &$0.0984\pm 0.0130~^{+0.0045}_{-0.0051}$&
    $0.1346\pm 0.0088~^{+0.0117}_{-0.0041}$\rbtrr\\\hline
  \end{tabular}\hftwo\vspace{0.5cm}
  \hftwo\begin{tabular}{|c||l|l|l|}
    \hline
    \multicolumn{1}{|c||}{$\rbthm\mean{Q}/\gev$}&
    \multicolumn{1}{|c|}{$\mean{\tau_C}$}&
    \multicolumn{1}{|c|}{$\mean{\rho}$}&
    \multicolumn{1}{|c|}{$\mean{C}$}
    \\\hline\hline
    $7.5$ &
    $0.1637\pm 0.0032~^{+0.0030}_{-0.0029}$&
    $0.1115\pm 0.0019~^{+0.0011}_{-0.0013}$&
    $0.5601\pm 0.0082~^{+0.0083}_{-0.0073}$
    \rbtrr\\\hline 
    $8.7$ &
    $0.1600\pm 0.0037~^{+0.0060}_{-0.0031}$&
    $0.1044\pm 0.0021~^{+0.0017}_{-0.0000}$&
    $0.5524\pm 0.0094~^{+0.0074}_{-0.0054}$
    \rbtrr\\\hline
    $15.0$&
    $0.1333\pm 0.0013~^{+0.0013}_{-0.0019}$&
    $0.0872\pm 0.0007~^{+0.0007}_{-0.0014}$&
    $0.4824\pm 0.0034~^{+0.0036}_{-0.0051}$
    \rbtrr\\\hline
    $17.8$&
    $0.1263\pm 0.0011~^{+0.0014}_{-0.0024}$&
    $0.0826\pm 0.0007~^{+0.0013}_{-0.0015}$&
    $0.4621\pm 0.0030~^{+0.0045}_{-0.0069}$
    \rbtrr\\\hline
    $23.6$&
    $0.1098\pm 0.0012~^{+0.0020}_{-0.0026}$&
    $0.0714\pm 0.0007~^{+0.0019}_{-0.0016}$&
    $0.4112\pm 0.0033~^{+0.0056}_{-0.0076}$
    \rbtrr\\\hline
    $36.7$&
    $0.0985\pm 0.0021~^{+0.0012}_{-0.0023}$&
    $0.0634\pm 0.0012~^{+0.0013}_{-0.0013}$&
    $0.3644\pm 0.0058~^{+0.0029}_{-0.0058}$
    \rbtrr\\\hline
    $57.9$&
    $0.0834\pm 0.0040~^{+0.0015}_{-0.0046}$&
    $0.0518\pm 0.0023~^{+0.0016}_{-0.0025}$&
    $0.3127\pm 0.0122~^{+0.0065}_{-0.0131}$
    \rbtrr\\\hline
    $81.3$&
    $0.0663\pm 0.0057~^{+0.0031}_{-0.0025}$&
    $0.0410\pm 0.0034~^{+0.0022}_{-0.0016}$&
    $0.2529\pm 0.0173~^{+0.0160}_{-0.0065}$\rbtrr\\\hline 
  \end{tabular}\hftwo\vspace{0.5cm}
  \hftwo\begin{tabular}{|c||l|l|}
    \hline
    \multicolumn{1}{|c||}{$\rbthm\mean{Q}/\gev$}&
    \multicolumn{1}{|c|}{$\mean{y_{fJ}}$}&
    \multicolumn{1}{|c|}{$\mean{y_{k_t}}$}
    \\\hline\hline
    $7.5$ &
    $0.1598\pm 0.0026~^{+0.0048}_{-0.0049}$&
    $0.3088\pm 0.0065~^{+0.0081}_{-0.0145}$
    \rbtrr\\\hline 
    $8.7$ &
    $0.1497\pm 0.0030~^{+0.0055}_{-0.0049}$&
    $0.2320\pm 0.0062~^{+0.0099}_{-0.0091}$
    \rbtrr\\\hline
    $15.0$&
    $0.1260\pm 0.0015~^{+0.0063}_{-0.0058}$&
    $0.1349\pm 0.0025~^{+0.0078}_{-0.0065}$
    \rbtrr\\\hline
    $17.8$&
    $0.1180\pm 0.0014~^{+0.0052}_{-0.0053}$&
    $0.1147\pm 0.0021~^{+0.0067}_{-0.0059}$
    \rbtrr\\\hline
    $23.6$&
    $0.1049\pm 0.0015~^{+0.0050}_{-0.0052}$&
    $0.0940\pm 0.0022~^{+0.0061}_{-0.0044}$
    \rbtrr\\\hline
    $36.7$&
    $0.0861\pm 0.0025~^{+0.0037}_{-0.0049}$&
    $0.0627\pm 0.0028~^{+0.0047}_{-0.0022}$
    \rbtrr\\\hline
    $57.9$&
    $0.0785\pm 0.0053~^{+0.0049}_{-0.0066}$&
    $0.0463\pm 0.0044~^{+0.0055}_{-0.0028}$
    \rbtrr\\\hline
    $81.3$&
    $0.0608\pm 0.0071~^{+0.0052}_{-0.0057}$&
    $0.0333\pm 0.0050~^{+0.0050}_{-0.0035}$\rbtrr\\\hline 
  \end{tabular}\hftwo
  \caption{Corrected mean values of the event shapes as a function of $Q$.
    The first uncertainty is statistical, the second systematic.}
  \label{tab:finalmeans}
\end{table}

%%% Local Variables: 
%%% mode: latex
%%% TeX-master: "draft"
%%% End: 
      % Measurement of Event Shapes
\section{Theoretical Framework}
\label{qcdframework}

Infrared and collinear safe event shape variables are presently
calculable in DIS up to next-to-leading order QCD\@. Several programs
are available.  However, for comparisons to real experimental
situations this is insufficient and a phenomenological description of
hadronization is needed.  Within the concept of power corrections one
assumes that a parameterization of the leading corrections to the perturbative
prediction can be obtained without modelling all the details of the
hadronization.
This leads to the notion of `universal' power corrections
with a definite $Q$ dependence, typically ${\cal O}(1/Q^p)$, given in
analytic form with a calculable coefficient for each event shape
observable.  The hope is that such a simplifying approach gives useful
insight in the interplay of perturbative and non-perturbative effects.

\subsection{pQCD Calculations}
\label{pqcdcalc}

\begin{figure} 
  \centering
  \includegraphics{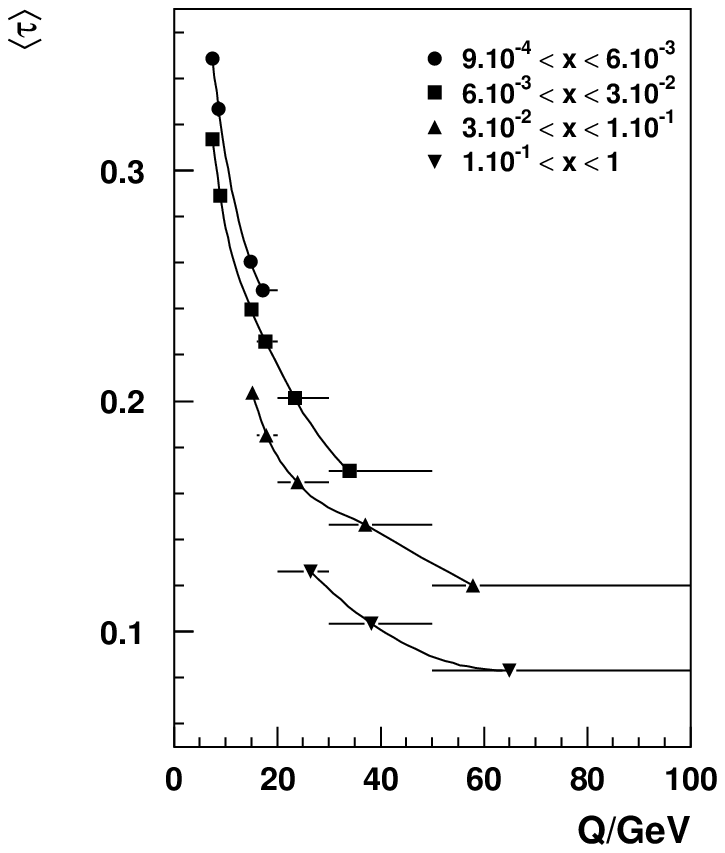}\hftwo%
  \includegraphics{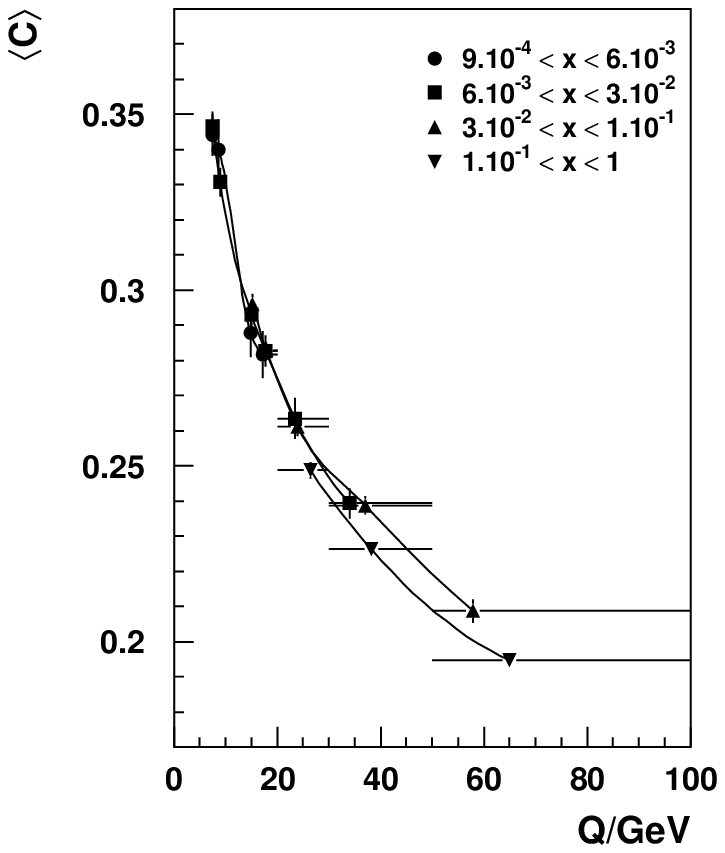}
  \caption {Mean values of $\tau$ (left) and $C$ (right) versus $Q$
    in four different bins of $x$ calculated with \DISENT\@.  The
    lines connect the means belonging to the same $x$ bin.}
  \label{fig:xdep}
\end{figure}

The mean value of an event shape variable $F$ can be written in second
order perturbative QCD as
\begin{eqnarray}
  \mean{F}^{{\rm pert}} & = &
  c_1(x,Q)\,\as(\mr) + \left[c_2(x,Q) + 
    \frac{\beta_0}{2\pi}\ln\frac{\mr}{Q}c_1(x,Q)\right]\as^2(\mr)\,,
  \label{eqn:fpert} 
\end{eqnarray}
where $\mr$ is the renormalization scale, $\beta_0 = 11 - 2/3\,N_{f}$
and $N_f = 5$ is the number of active flavours.  In contrast to \ee
annihilation, where the coefficients $c_1$ and $c_2$ are constant,
in DIS they depend on $x$ according to the parton density functions
and the accessible $x$-range at different values of $Q$ (see
figure~\ref{fig:kinescat}).  Examples for a strong and a weak variation
with $x$ of the mean values $\mean{\tau}$ and $\mean{C}$ versus $Q$
are presented in figure~\ref{fig:xdep}.
Average values of the coefficients $c_{1}$ and $c_{2}$ are calculated
separately for every $Q$ bin with its specific range of $x$-values.
This approximation by step functions is the origin of the steps
exhibited by the curves in
figures~\ref{fig:DWfit1} and~\ref{fig:DWfit2}. The slope within one
bin is due to the variation of $\as$ with $Q$.  In the previous
publication~\cite{h1eventshapes}, average coefficients determined for
the complete range in $Q$ have been employed.

The event shape means are evaluated at the scale $\mr = Q$ with the
\DISENT~\cite{disent} program which treats deep-inelastic $ep$
scattering to ${\cal O}(\as^2)$ in the $\overline{\mbox{MS}}$ scheme
and employs the subtraction method for the necessary integrations.  In
order to test the reliability of the perturbative predictions, careful
comparisons of \MEPJET~\cite{mepjet}, \DISENT\ and
\DISASTER++~\cite{disaster} have been carried out~\cite{nlocheck}.
Due to intrinsic restrictions of the integration technique applied in
\MEPJET\ (phase space slicing), however, it can not be used for the
mean event shapes.  In general, good agreement at the percent level is
observed but some discrepancies have been revealed. These are
partially understood and an improved \DISENT\ version has been
employed.  The effects are hardly visible in the event shape spectra,
but the mean values have increased considerably in the low $Q$ region
compared to the calculations used in~\cite{h1eventshapes}.  The worst
case is the jet broadening $B$ whose mean value at $Q = 7.5\gev$ rose
by about $14\%$,
being still slightly different from the \DISASTER++ calculation.%
\footnote{Artificially increasing the \DISENT\ predictions for the two
  low $Q$ means of the jet broadening by $10\%$ and $5\%$ respectively
  improves the consistency of the fits described in
  section~\ref{dwfits}, especially with respect to $\as$.}
  
The parton density distributions MRSA'~\cite{mrsap} with $\asmz =
0.115$ are used as standard.  Other sets, i.e.~MRST99~\cite{mrst99}
and CTEQ4~\cite{cteq}, are investigated as well to estimate systematic
uncertainties.

\subsection{Power Corrections}
\label{powercorrections}

Hadronization effects on event shapes are treated within the concept
of power corrections~\cite{dwmodel,webber}. The observable mean values
can be written as
\begin{eqnarray}
  \fmean & = & \fpert + \fpow\,,
  \label{eqn:ftot}
\end{eqnarray}
with $\fpert$ given by eq.~(\ref{eqn:fpert}). The hadronization
contributions $\fpow$ are expected to be proportional to $1/Q^p$ with
exponents $p = 1$ or $p = 2$ depending on the observable. Two types of
parameterizations for $\fpow$ will be investigated.

In a simplistic approach inspired by the longitudinal phase space or
tube model, described e.g.~in~\cite{esw}, one has
\begin{eqnarray}
  \fpow = \frac{\lambda_{1,\,F}}{Q} & {\rm or} &
  \fpow = \frac{\lambda_{2,\,F}}{Q^2}\,,
  \label{eqn:1/Q}
\end{eqnarray}
where $\lambda_{1,\,F}$ and $\lambda_{2,\,F}$ are constants.  One
expects $\lambda_1$ to be in the order of $\simeq 1\gev$ and terms
with exponents $p>1$ to be negligible except for the differential
two-jet rate $y_{k_t}$.  Even within this simple model it is possible
to derive approximate relations between the constants $\lambda_1$ for
different event shapes, e.g.~$\lambda_{1,\,\tau}\simeq
2\,\lambda_{1,\,\rho}$, consistent with $e^+e^-$ data~\cite{esw}.

In the model pioneered by Dokshitzer and Webber~\cite{dwmodel,webber}
the idea is to attribute $1/Q^p$ power-law corrections to soft gluon
phenomena associated with the behaviour of the strong coupling at
small scales.  This leads to the notion of a universal infrared-finite
effective coupling $\aeff(\mr)$ which replaces the perturbative form
in the infrared region $\mr < \mi$ where $\mi$, the {\em infrared
  matching}\/ scale, has to fulfil $\Lambda_{\rm QCD} \ll \mi \approx
2\gev \ll Q$.  At the expense of one new non-perturbative parameter
$\ap(\mi)$, corresponding to the $(p-1)$th moment of the effective
coupling $\aeff(\mr)$ when integrated from $0$ up to $\mi$, the power
corrections to all event shapes with the same exponent $p$ can be
related via~\cite{fpowd,fpoww}
\begin{eqnarray}
  \fpow & = & a_F\,{\cal P}\,, 
  \label{fpow}  
\end{eqnarray}
\begin{eqnarray}
  {\cal P} & = & \frac{4C_F}{\pi\ p}{\cal M'}
  \left(\frac{\mi}{\mr}\right)^p
  \left [\ap(\mi) - \as(\mr) - \frac{\beta_0}{2\pi}
    \left(\ln\frac{\mr}{\mi} + \frac{K}{\beta_0} + \frac{1}{p}
    \right)\as^2(\mr)\right]
  \label{eqn:calp}
\end{eqnarray}
where $C_F = 4/3$, $K = 67/6 - \pi^2/2 - 5/9\,N_f$ and $N_f = 5$ as in
the perturbative part. Again, $\mr$ is identified with $Q$ in this
study.  The subtractions proportional to $\as$ and $\as^2$ serve to
avoid double counting.

The coefficients $a_F$, given in table~\ref{tab:2parDWfit}, depend on
the observable $F$ but can in principle be derived from a perturbative
ansatz, although not yet available for the variable $y_{k_t}$.  Some
of them have changed considerably --- e.g.~$a_B$ and $a_C$ by factors
of $4$ and $2$ respectively --- from their original
values~\cite{fpoww} applied in previous
investigations~\cite{h1eventshapes,humdis98}.  Ambiguities in
calculating the $a_F$ predictions could be resolved with the advent of
a two-loop analysis provided an additional common coefficient, the
{\em Milan factor}\/ ${\cal M}$~\cite{fpowd,milan2000} with
\begin{eqnarray}
  {\cal M'} & = & \frac{2}{\pi}{\cal M} =\frac{2}{\pi}
  \left(1 + \frac{4.725 - 0.104\, N_f}{\beta_0} \right)
  \label{eqn:milan}
\end{eqnarray}
is applied.  The numerical value is ${\cal M'} \simeq 0.95$ for
$N_f=3$ flavours relevant for gluon radiation at low scales.%
\footnote{Note that the original derivation, which lead to
  ${\cal M'} \simeq 1.14$, has recently been corrected~\cite{milan2000}.}

For all event shapes under study a power suppression exponent of $p =
1$ is expected except for the two-jet rate $y_{k_t}$ where $p = 2$.
The power correction parameter $\an$ is estimated to be $\simeq 0.5$
whilst $\ao$ is essentially unknown.

The jet broadening $B$ is expected to behave differently from the
other event shapes and eq.~(\ref{eqn:calp}) should contain an
additional enhancement.  The originally proposed factor of
$\ln(Q/Q_0)$ \cite{fpoww}, with $Q_0 \sim {\cal O}(\mi)$ an unknown
scale, could not be supported by the H1 data~\cite{h1eventshapes}.
This observation stimulated a theoretical reexamination leading to the
following power correction for the jet broadening~\cite{bpow}
\begin{eqnarray}
  \mean{B}^{\rm pow} & = & a_B\,a_B'\,{\cal P} =
  a_B \left(\frac{\pi}{2\sqrt{2C_F\as(1+K/(2\pi)\cdot\as)}} + \frac{3}{4}
    - \frac{\beta_0}{12C_F} + \eta_0 \right){\cal P}\,,
  \label{eqn:bpow}
\end{eqnarray}
where $a_B = 1/2$, $\eta_0 = -0.614$ and $\as$ has to be evaluated at
the scale $e^{-\frac{3}{4}}\cdot\mr$.  The enhancement term $a_B'$ is
substantial with a slow variation of $1.6-2.2$ over the measured $Q$
range.  Strictly, this formula has been derived for \ee annihilation,
but should be applicable for DIS as well. The accuracy of the
coefficient, however, is only of the order of $1$~\cite{bpow}!  All
coefficients $a_F$ are considered to be input parameters and
systematic uncertainties do not account for approximations, e.g.\ the
neglect of quark mass effects, inherent in their derivation.

%%% Local Variables: 
%%% mode: latex
%%% TeX-master: "draft"
%%% End: 
     % Theoretical Framework
\section{QCD Analysis of Event Shape Means}
\label{meanfits}

In order to get an impression of the impact of hadronization,
figures~\ref{fig:dndFhl1} and~\ref{fig:dndFhl2} show the experimental
event shape spectra in comparison with calculations of perturbative
QCD\@. At high values of $Q$ the effects are small, as expected, and
the unfolded and partonic spectra approach each other.  At low $Q$,
data and calculations look very different and non-perturbative effects
become prominent.  It is particularly interesting to note that both
two-jet rates $y_{fJ}$ and $y_{k_t}$ already exhibit small hadronization
corrections at modest momentum transfer, a characteristic very
different from the other variables.

In this section a comprehensive study of event shape means will be
presented in order to pin down the analytical form and magnitude of
power-law corrections and to test their universal nature.  The
strategy is to investigate the $Q$ dependence of $\fmean$ as given in
table~\ref{tab:finalmeans} assuming the ansatz of
eq.~(\ref{eqn:ftot}).  The perturbative part is given by the QCD
expression of eq.~(\ref{eqn:fpert}).  Two variants of power
corrections $\fpow$ will be tested: the tube model and the approach
pioneered by Dokshitzer and Webber.

\subsection{{\boldmath $1/Q^p$} Fits}
\label{qpfits}

Keeping $\asmz$ fixed to $0.119$ and
parameterizing hadronization contributions by eq.~(\ref{eqn:1/Q})
yields the fit results for $\lambda_{1,\,F}$ and $\lambda_{2,\,F}$
given in table~\ref{tab:1parmufit}.  With the exception of $\tau$, $B$
and $y_{fJ}$ acceptable results can not be achieved in the case of the
$\lambda_1/Q$ term. Note that for $\tau$ and $B$ $\lambda_1$ complies
with the expectations.  Applying $\lambda_2/Q^2$ corrections the
$\chi^2$ values worsen dramatically and lead to the rejection of this ansatz.

\begin{table}[htb]
  \centering
  \begin{tabular}{|c||r|r||r|r||r|r|}
    \hline
    \multicolumn{1}{|c||}{\rbthm$\fmean$} &
    \multicolumn{1}{|c|}{$\lambda_{1,\,F}/\gev$} & 
    \multicolumn{1}{|c||}{$\chin$} & 
    \multicolumn{1}{|c|}{$\lambda_{2,\,F}/\gevq$} & 
    \multicolumn{1}{|c||}{$\chin$} &
    \multicolumn{1}{|c|}{$\asmz$} & 
    \multicolumn{1}{|c|}{$\chin$} \\\hline\hline
    $\mean{\tau}$ &
    $0.71 \pm 0.03$  & $  2$ &
    $7.2  \pm 0.3$   & $ 23$ &
    $0.131\pm 0.001$ & $  1$ \rbtrr\\\hline
    $\mean{B}$ &
    $0.55 \pm 0.02$  & $  1$ &
    $5.2  \pm 0.2$   & $ 28$ &
    $0.134\pm 0.001$ & $ 24$ \rbtrr\\\hline
    $\mean{\tau_C}$ &
    $0.73 \pm 0.01$  & $ 29$ &
    $7.1  \pm 0.2$   & $212$ &
    $0.152\pm 0.001$ & $  2$ \rbtrr\\\hline
    $\mean{\rho}$ &
    $0.54 \pm 0.01$  & $ 38$ &
    $5.2  \pm 0.1$   & $283$ &
    $0.160\pm 0.001$ & $  1$ \rbtrr\\\hline
    $\mean{C}$ &
    $2.33 \pm 0.03$  & $ 66$ &
    $20.8 \pm 0.4$   & $327$ &
    $0.148\pm 0.001$ & $  5$ \rbtrr\\\hline
    $\mean{y_{fJ}}$ &
    $-0.13\pm 0.01$  & $  2$ &
    $-1.2 \pm 0.1$   & $  7$ &
    $0.113\pm 0.001$ & $  1$ \rbtrr\\\hline
    $\mean{y_{k_t}}$ &
    $-0.32\pm 0.02$  & $  7$ &
    $-3.1 \pm 0.3$   & $ 21$ &
    $0.110\pm 0.001$ & $  9$ \rbtrr\\\hline
  \end{tabular}
  \caption{Results of fits to event shape means.
    Left: fit of $\lambda_{1,\,F}$ with $\asmz = 0.119$,
    center: fit of $\lambda_{2,\,F}$ with $\asmz = 0.119$ and
    right: fit of $\asmz$. Uncertainties are statistical only.}
  \label{tab:1parmufit}
\end{table}

Allowing for test purposes a variation of $\asmz$ in the perturbative
expression while neglecting power corrections, some of these fits
actually work but result in large discrepancies of the strong coupling
with respect to a world average of $\asmz = 0.119 \pm
0.002$~\cite{pdg}.  Neither a power-like correction according to
eq.~(\ref{eqn:1/Q}) nor pure pQCD, exploiting the logarithmic $Q$
dependence of the strong coupling, are sufficient to describe all
data.  However, fitting $\asmz$ and $\lambda_{1,\,F}$ or
$\lambda_{2,\,F}$ simultaneously, does not lead to satisfactory
results either.  Due to an extreme anti-correlation between the two
parameters there is a tendency to minimize the power contribution at
the cost of unphysical shifts of $\asmz$.  Only $B$ and $y_{fJ}$
produce correlated but reasonable numbers. Details can be found
in~\cite{rabbertz}.  These studies suggest that some form of
combined power-like and logarithmic $Q$ dependence is needed to account
for the observed medium ($\tau$, $B$), large ($\tau_C$, $\rho$, $C$) and
small ($y_{k_t}$, $y_{fJ}$) hadronization corrections.

\subsection{Fits in the approach initiated by Dokshitzer and Webber}
\label{dwfits}

\begin{figure} 
%  \centering
  \hftwo\includegraphics{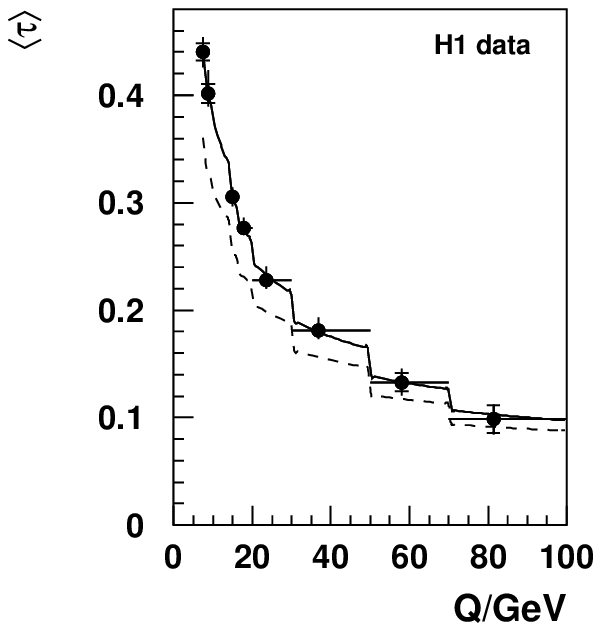}\hftwo%
  \includegraphics{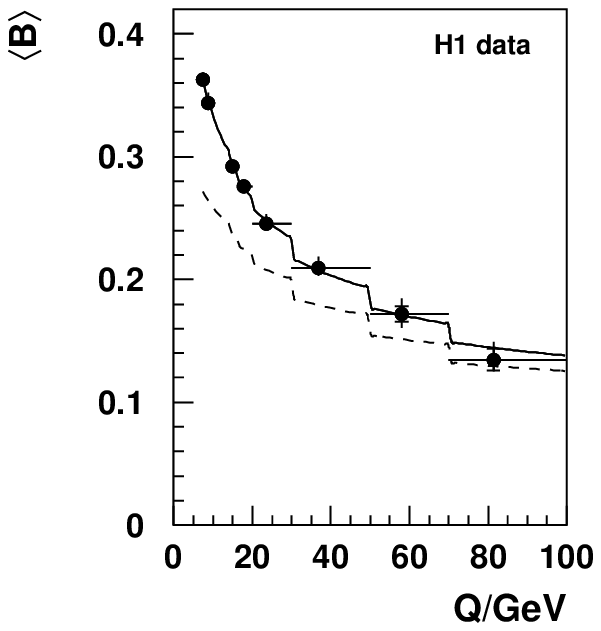}\hftwo\\
  \hftwo\includegraphics{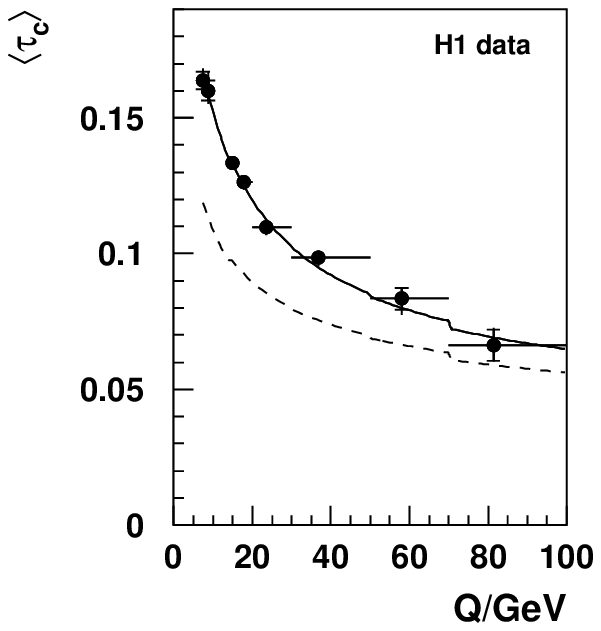}\hftwo%
  \includegraphics{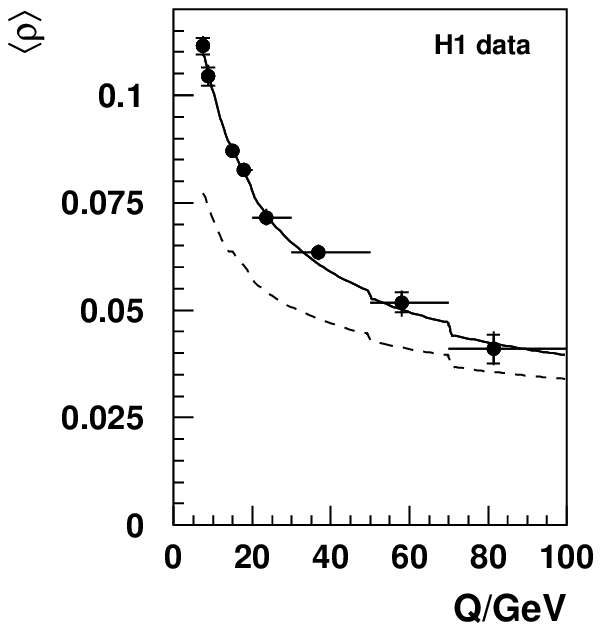}\hftwo\\
  \hftwo\includegraphics{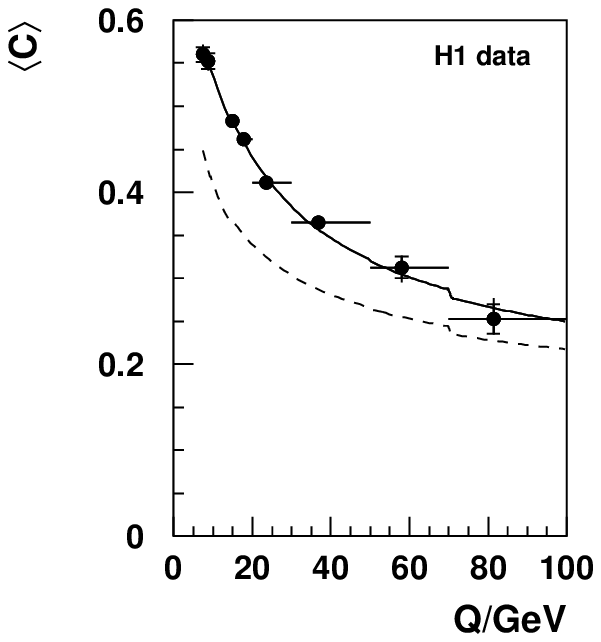}\hftwo%
  \includegraphics{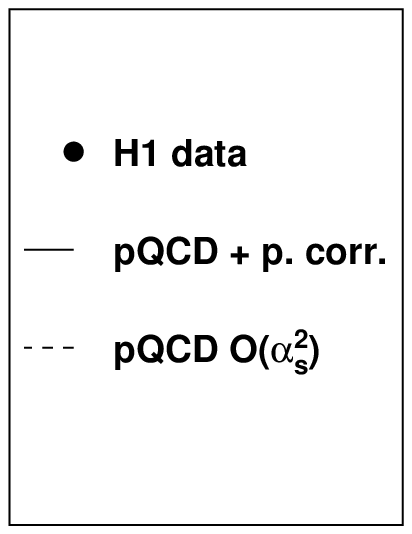}\hftwo
  \caption{Mean values (full symbols) of $\tau$, $B$, $\tau_C$,
    $\rho$ and $C$ as a function of $Q$.  The error bars represent
    statistical and systematic uncertainties.  The full line
    corresponds to a power correction fit according to the
    Dokshitzer--Webber approach.  The dashed line shows the pQCD
    contribution of \DISENT\ in these fits.}
  \label{fig:DWfit1}
\end{figure}

\begin{figure} 
  \centering
  \includegraphics{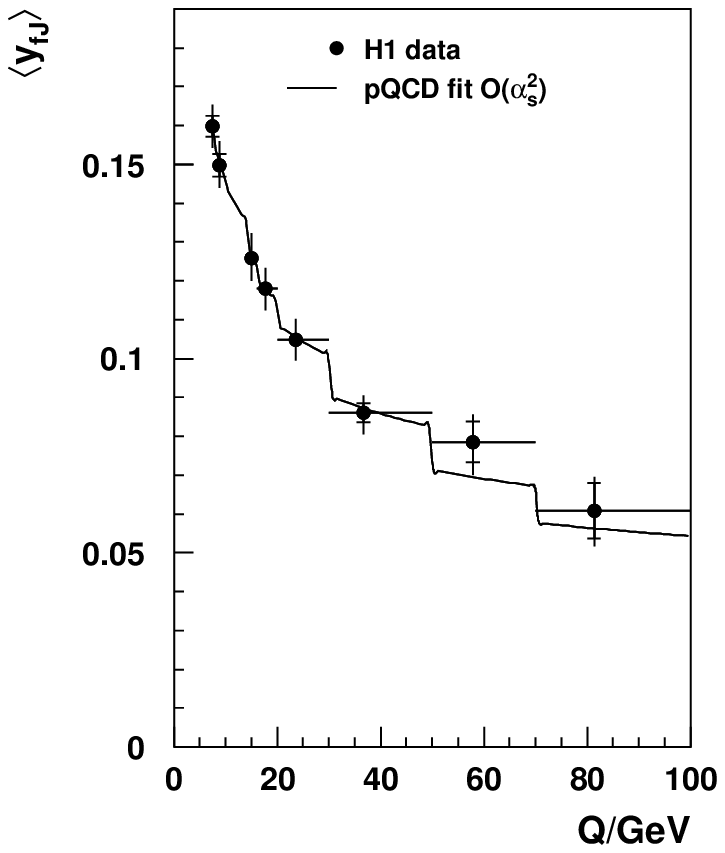}\hftwo%
  \includegraphics{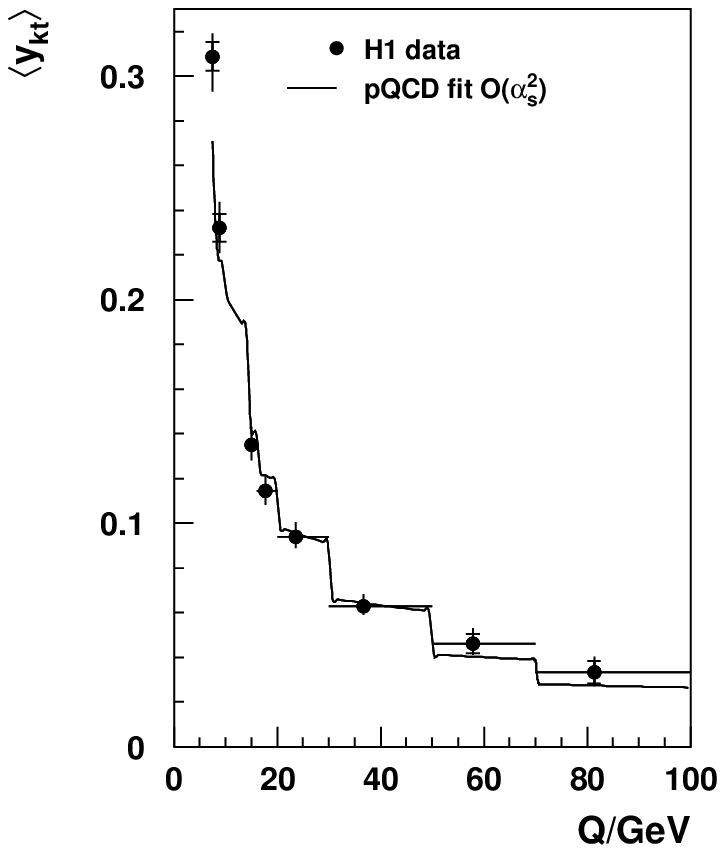}
  \includegraphics{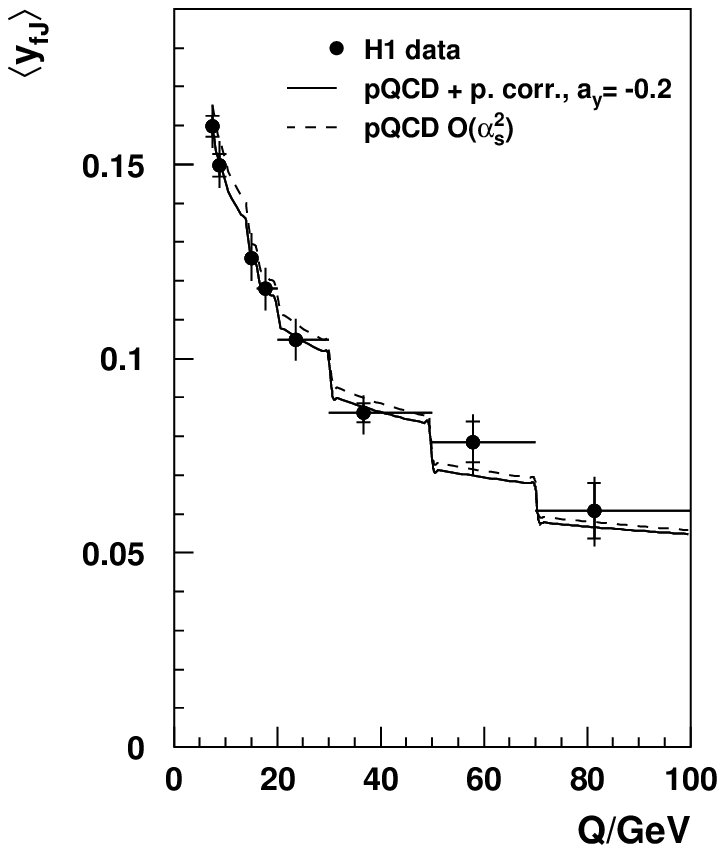}\hftwo%
  \includegraphics{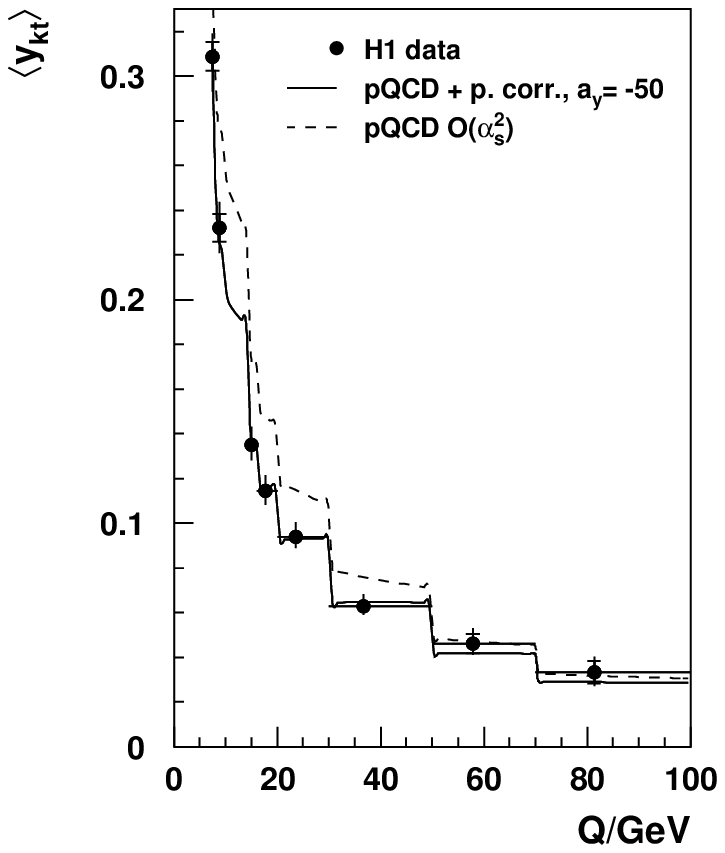}
  \caption{Mean values (full symbols) of $y_{fJ}$
    and $y_{k_t}$ as a function of $Q$.  The error bars represent
    statistical and systematic uncertainties.  Upper part: The full
    line corresponds to a fit of the pQCD calculation without power
    contribution.  Lower part: The full line corresponds to a power
    correction fit according to the Dokshitzer--Webber approach.  The
    dashed line shows the pQCD contribution of \DISENT\ in these fits.}
  \label{fig:DWfit2}
\end{figure}

Fits in the approach initiated by Dokshitzer and Webber are performed
separately for each event shape.  The data are very well described by
pQCD plus these analytical power corrections
as shown in figures~\ref{fig:DWfit1} and~\ref{fig:DWfit2}.%
\footnote{The steps are due to the $x$ dependence of the pQCD
  calculations (see section~\ref{pqcdcalc}).}
The fit results for the power correction parameters $\ap$ and the
strong coupling $\asmz$ are compiled in table~\ref{tab:2parDWfit}.
For a discussion of the quoted uncertainties see
section~\ref{syserrors}.  Theoretical systematics for the fit
parameters of the differential two-jet rates $y_{fJ}$ and $y_{k_t}$
are not given due to the essentially unknown coefficients $a_F$ in the power
correction predictions.

\begin{table}
  \centering
  \begin{tabular}{|c|c|c||c|c|r|r|}
    \hline
    \multicolumn{1}{|c|}{\rbthm$\fmean$} &
    \multicolumn{1}{|c|}{$a_F$} & 
    \multicolumn{1}{|c||}{$p$} & 
    \multicolumn{1}{|c|}{$\ap(\mi=2\gev)$} & 
    \multicolumn{1}{|c|}{$\asmz$} & 
    \multicolumn{1}{|c|}{$\chin$} &
    \multicolumn{1}{|c|}{$\kappa/\%$} \\\hline\hline
    $\mean{\tau}$ & $1$ & $1$ &
    $0.503~^{+0.043}_{-0.053}~^{+0.053}_{-0.068}$ &
    $0.1190~^{+0.0075}_{-0.0054}~^{+0.0073}_{-0.0069}$ &
    $0.5$ & $-98$ \rbtrr\\\hline
    $\mean{B}$ & $1/2\cdot a'_{B}$ & $1$ &
    $0.537~^{+0.017}_{-0.012}~^{+0.028}_{-0.039}$ &
    $0.1113~^{+0.0036}_{-0.0028}~^{+0.0049}_{-0.0051}$ &
    $0.7$ & $-69$ \rbtrr\\\hline
    $\mean{\rho}$ & $1/2$ & $1$ &
    $0.597~^{+0.009}_{-0.010}~^{+0.050}_{-0.057}$ &
    $0.1374~^{+0.0024}_{-0.0032}~^{+0.0110}_{-0.0096}$ &
    $1.1$ & $-32$ \rbtrr\\\hline
    $\mean{\tau_C}$ & $1$ & $1$ &
    $0.503~^{+0.008}_{-0.010}~^{+0.043}_{-0.047}$ &
    $0.1310~^{+0.0023}_{-0.0028}~^{+0.0098}_{-0.0089}$ &
    $1.2$ & $-22$ \rbtrr\\\hline
    $\mean{C}$ & $3\pi/2$ & $1$ &
    $0.447~^{+0.005}_{-0.007}~^{+0.032}_{-0.038}$ &
    $0.1301~^{+0.0016}_{-0.0020}~^{+0.0103}_{-0.0091}$ &
    $0.8$ & $+36$ \rbtrr\\\hline\hline
    $\mean{y_{fJ}}$ & $1$ & $1$ &
    $0.28~^{+0.02}_{-0.02}$ &
    $0.105~^{+0.005}_{-0.006}$ &
    $0.8$ & $-72$ \rbtrr\\\hline
    $\mean{y_{fJ}}$ & $-0.2^*$ & $1$ &
    $0.37~^{+0.20}_{-0.21}$ &
    $0.116~^{+0.008}_{-0.009}$ &
    $0.6$ & $+98$ \rbtrr\\\hline\hline
    $\mean{y_{k_t}}$ & $1^\dagger$ & $1^\dagger$ &
    $0.65~^{+0.03}_{-0.04}$ &
    $0.001~^{+0.022}_{-0.012}$ &
    $7.2$ & $-98$ \rbtrr\\\hline
    $\mean{y_{k_t}}$ & $1^\dagger$ & $2$ &
    $1.50~^{+0.23}_{-0.39}$ &
    $0.099~^{+0.007}_{-0.005}$ &
    $3.6$ & $-92$ \rbtrr\\\hline
    $\mean{y_{k_t}}$ & $-50^*$ & $2$ &
    $0.34~^{+0.12}_{-0.11}$ &
    $0.124~^{+0.015}_{-0.014}$ &
    $0.6$ & $+99$ \rbtrr\\\hline
  \end{tabular}
  \caption{Results of fits \`{a} la Dokshitzer--Webber 
    for the event shape means. The coefficients $a_F$ and exponents
    $p$ of the power corrections are given as well.
    The first uncertainty contains statistics and experimental systematics,
    the second is an estimate of theoretical uncertainties
    (omitted for $y$ variables).
    $\kappa$ denotes the correlation coefficient
    between $\ap$ and $\asmz$.
    The starred coefficients of the $y$ variables are derived
    from a fit procedure, whereas the coefficients marked
    with $^\dagger$ are trials/guesses.}
  \label{tab:2parDWfit}
\end{table}

First the fits to event shape variables defined in the current
hemisphere will be discussed, i.e.\ not including the two-jet
variables $y_{fJ}$ and $y_{k_t}$.
With the exception of $\tau$, correlations are reduced
compared with the tube model and the $\chi^2$ values are reasonable.
It is interesting to note that the new calculations of the power
corrections for the jet broadening, eqs.~(\ref{fpow})
and~(\ref{eqn:bpow}), are now able to describe the data well.
Neglecting the enhancement factor gives fit results of
$\an = 0.661~^{+0.024}_{-0.021}~^{+0.028}_{-0.039}$ and
$\asmz = 0.1169~^{+0.0036}_{-0.0027}~^{+0.0049}_{-0.0051}$ with $\chin = 0.8$.

\begin{figure} 
  \centering \includegraphics[height=13.cm]{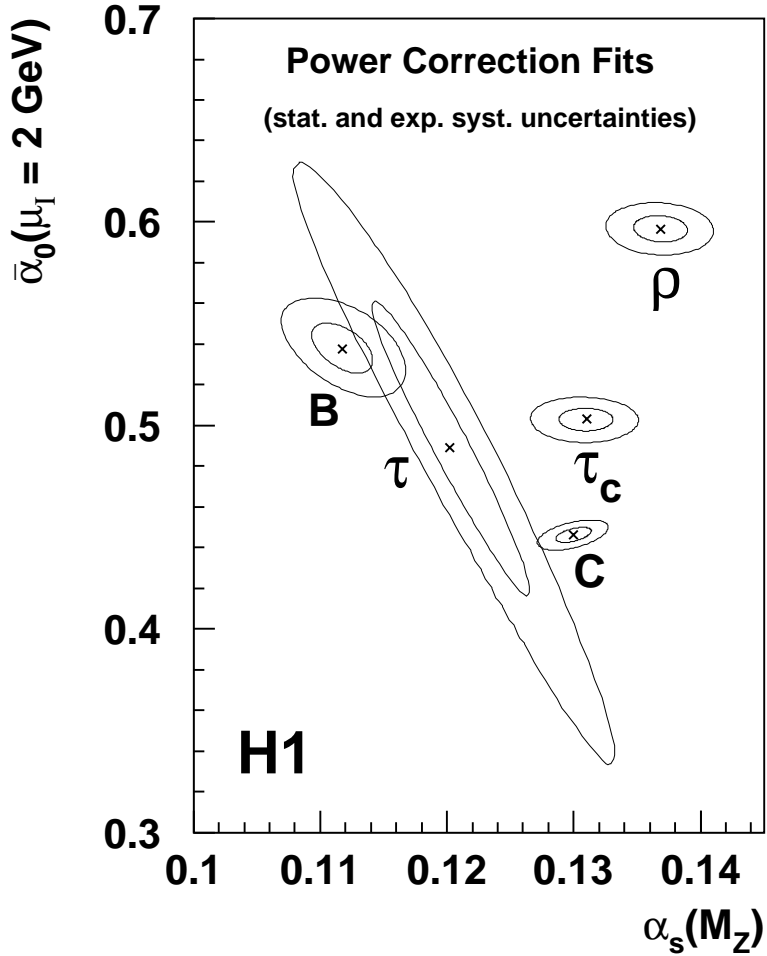}
  \caption{Results of the power correction fits to the mean values of $\tau$,
    $B$, $\rho$, $\tau_C$ and $C$ with the contours of
    $\chi^2(\as,\an)=\chi^2_{\rm min}+1$ and
    $\chi^2(\as,\an)=\chi^2_{\rm min}+4$
    including statistical and experimental systematic uncertainties.}
  \label{fig:ellipses}
\end{figure}

The fitted parameters are displayed in the $\as - \an$ plane of
figure~\ref{fig:ellipses}. Note that the experimental uncertainties,
statistics and systematics, are in general smaller than the
theoretical uncertainties of $5-10\%$.  The parameters $\an$ scatter
around the expectation of $\an \approx 0.5$ within about $20\%$ and
are compatible with the assumption of universality
in the Dokshitzer--Webber approach.  The spread of the strong coupling
constant appears to be uncomfortably large and one observes a group of
higher $\asmz$ values for those variables which do not make use of the
boson axis.  Possible explanations are missing higher order QCD
corrections, expected to be different for each event shape variable,
and/or incomplete knowledge of the power correction coefficients.
Allowing the coefficients $a_F$ to vary by arbitrary factors of $2$
and $1/2$ in order to study the effect on $\an$ and $\asmz$, one
observes shifts of the ellipses of figure~\ref{fig:ellipses}
approximately along the main diagonal.  Larger coefficients $a_F$
induce smaller values of $\an$ and $\asmz$ and vice versa.

As already seen in section~\ref{qpfits} both two-jet rates $y$ defined
in the whole phase space of the Breit system exhibit only small
hadronization effects, much smaller than e.g.\ thrust.  For the
two-jet rate $y_{fJ}$ based on the \JADE\ algorithm an exponent $p=1$
and a coefficient $a_{y_{fJ}} = 1$ for the power corrections have been
given in~\cite{webber}.  Formally, one obtains an acceptable fit but
rather low and unreasonable numbers for $\an$ and $\asmz$.  In
particular any value of $\an \lesssim 0.3$ makes no sense within the
concept of this model.  This leads to the conclusion that the power
correction coefficient of ref.~\cite{webber}, which has not been
reinvestigated since, may not be correct.  Instead, more consistent
results can be obtained by either neglecting hadronization corrections
altogether or by taking a small, negative coefficient of
$a_{y_{fJ}}=-0.2$ (see figure~\ref{fig:DWfit2} and
table~\ref{tab:2parDWfit}).\pagebreak

In contrast to the other event shapes the hadronization correction to
$\mean{y_{k_t}}$ is expected to decrease with a larger power of $p =
2$~\cite{webber}.  In fact a power law behaviour with $p = 1$ and
$a_{y_{k_t}} = 1$ is strongly disfavoured by the data.  Other choices
of $a_{y_{k_t}}$ can not accommodate for reasonable values of $\an
\gtrsim 0.3$.  For a power exponent of $p=2$ the coefficient
$a_{y_{k_t}}$ is basically unknown and has to be determined in
addition to $\ao$. One can try to fit simultaneously $\ao$, $\asmz$
and $a_{y_{k_t}}$.  This set of parameters, however, is strongly
correlated. Nevertheless, the fit converges properly giving
$\ao = 0.34\pm 0.05$,
$\asmz = 0.125\pm 0.006$ and $a_{y_{k_t}}=-52\pm 3$ with $\chin =
0.6$ and statistical uncertainties only.
Reinserting a value of $a_{y_{k_t}} = -50$ gives the entry in
table~\ref{tab:2parDWfit}.  To prove a power correction term
$\propto 1/Q^2$, however, is currently not possible because the
observed hadronization effects are small.  The two-jet rate data
$y_{k_t}$ are certainly consistent with quadratic power law
corrections, but more experimental precision and theoretical input is
needed.  Figure~\ref{fig:DWfit2} shows fits to the mean values without
and with hadronization contributions.

In summary, the experimental observations are very well described
within the approach initiated by Dokshitzer and Webber.
The analytical form of the power correction contributions appears to
be adequate. Also the magnitude is of the right order for those event
shape variables where updated calculations exist, supporting the
notion of approximate universal power corrections.

\subsection{Systematic Uncertainties}
\label{syserrors}

The procedure to estimate the systematic uncertainties in
table~\ref{tab:2parDWfit} is to repeat the fits under variation of
every prominent systematic effect. The discrepancy compared to the
standard result is attributed to a corresponding uncertainty.  In case
of deviations in the same direction for the variation of one primary
source, e.g.~an upwards and downwards modification of an energy scale,
only the larger one is considered for the evaluation of the total
uncertainty.  The latter is derived from all contributions added in
quadrature.  An exception is the unfolding procedure whose influence
is estimated as explained in section~\ref{finalmeans}.  The following
systematic effects are investigated:\vspace{-0.25cm}
\begin{itemize}
\item Experimental uncertainties
  \begin{enumerate}
  \item Usage of four correction procedures
  \item Variation of the electromagnetic energy scale of the
    calorimeters by $\pm (1\% - 3\%)$
  \item Variation of the hadronic energy scale of the LAr calorimeter
    by $\pm 4\%$
  \end{enumerate}
\item Theoretical uncertainties
  \begin{enumerate}
  \item Variation of the renormalization scale $\mr^2$ by factors of
    $2$ and $1/2$
  \item Variation of the factorization scale $\mf^2$ by factors of $4$
    and $1/4$
  \item Variation of the infrared matching scale $\mi$ by $\pm
    0.5~\gev$
  \item Usage of MRST99 parton density functions~\cite{mrst99} with
    larger and smaller gluon contributions and usage of MRSA' parton
    density functions~\cite{mrsap} with strong couplings $\asmz=0.105$
    up to $0.130$, different from the standard set
  \item Usage of different parton density functions
    CTEQ4A2~\cite{cteq} with similar $\asmz$
  \end{enumerate}
\end{itemize}

The experimental sources are already discussed in
section~\ref{finalmeans}. The renormalization scale $\mr$ and the
factorization scale $\mf$ are arbitrary since in a complete theory the
calculations do not depend on any specific choice.  But in reality one
has only an approximate theory yielding residual dependences due to
neglected higher orders.  To avoid the appearance of large logarithms
in the computations it is recommended to identify the scales with a
process relevant scale chosen to be $Q$ in the present analysis.  To
estimate the effect of higher orders it is {\em conventional}\/ to
vary $\mr^2$ and $\mf^2$ by an arbitrary factor of $4$.  In the
case of $\mr^2$ one has to reduce this factor to $2$ because of the
infrared matching condition $\Lambda_{\rm QCD} \ll \mi \approx 2~\gev
\ll \mr$.

The variation of the infrared matching scale by $\mi = (2 \pm
0.5)\gev$ follows the original proposal~\cite{webber}.  Note that this
affects only the $\asmz$ uncertainty.  The parameter $\an$ explicitly
depends on $\mi$.

The last two points account for uncertainties in the gluon content of
the proton and the fact that $\asmz$ has implicitly already been used
in deriving the parton density functions which may bias the
computations.  The same is true for the choice of a parameterization
for the parton density functions.  Therefore five alternative sets,
two each with different gluons or $\asmz$ respectively and one with
approximately the same $\asmz$ but another parameterization, are
chosen for a reevaluation of the \DISENT\ calculations.

\begin{figure}
%  \centering
  \hftwo\includegraphics[height=7.5cm]{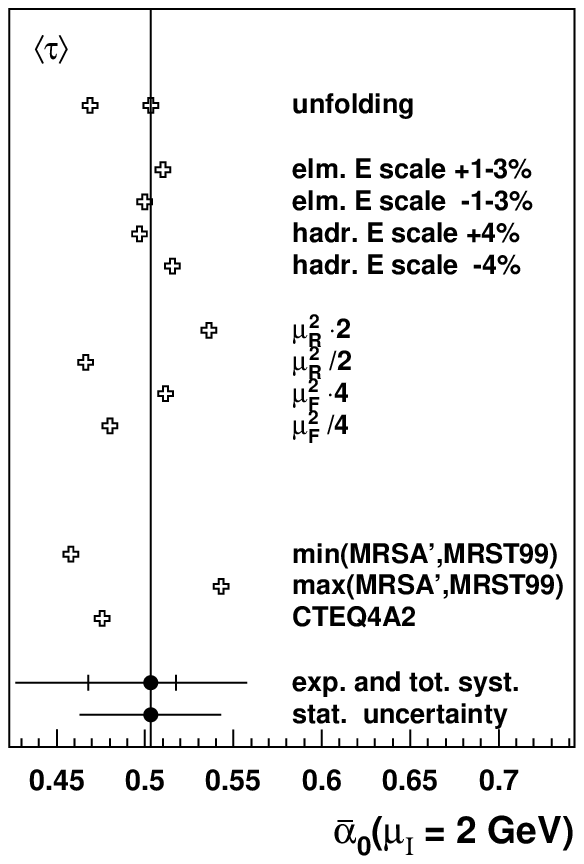}\hftwo%
  \includegraphics[height=7.5cm]{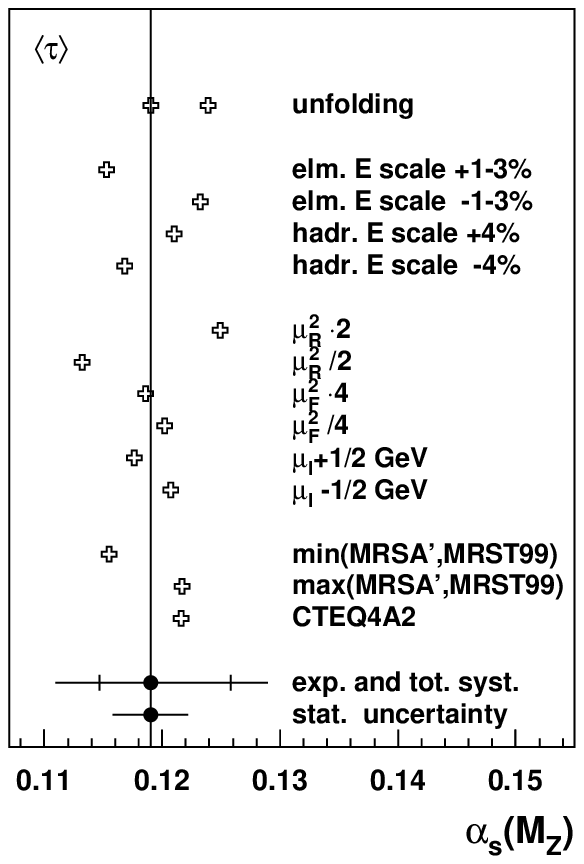}\hftwo\\
  \hftwo\includegraphics[height=7.5cm]{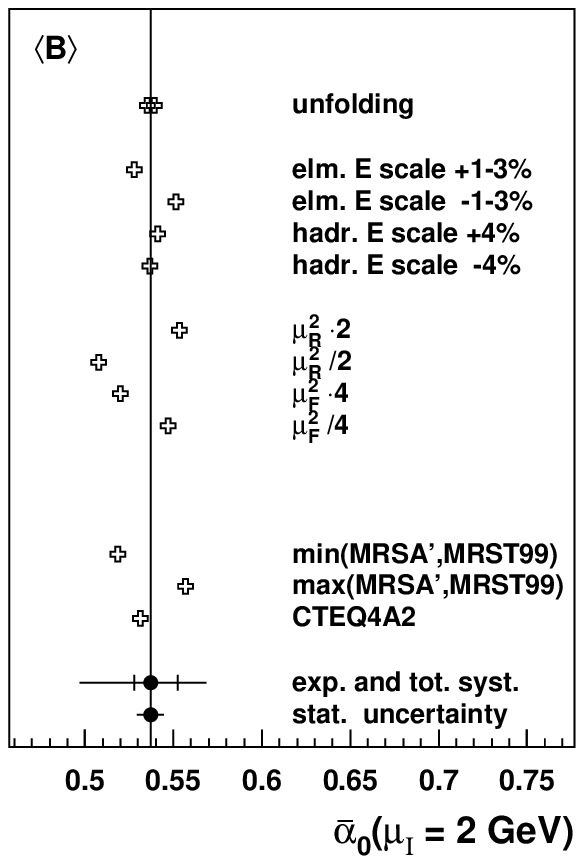}\hftwo%
  \includegraphics[height=7.5cm]{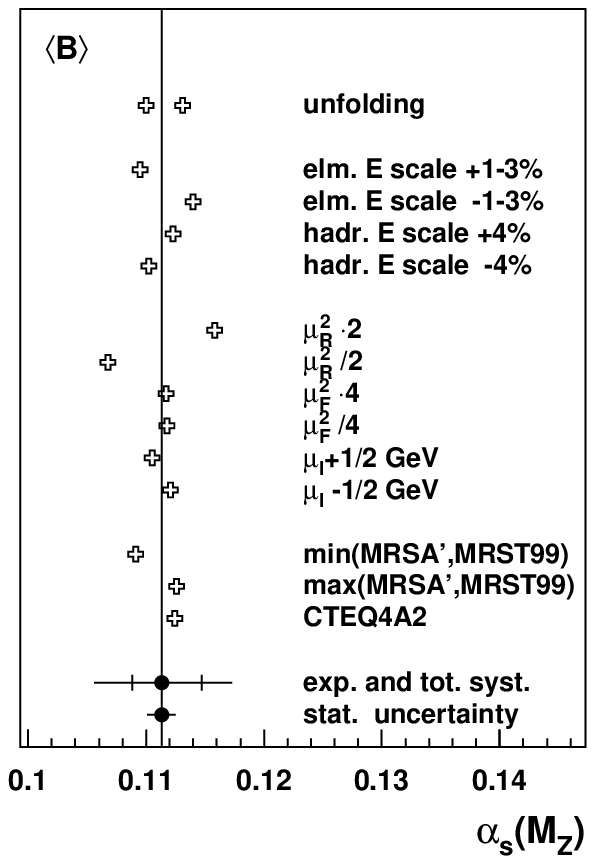}\hftwo
  \caption{Systematic uncertainties of $\anmi$  and $\asmz$ 
    for $\tau$ (top) and $B$ (bottom).}
  \label{fig:syserr1}
\end{figure}

\begin{figure}
%  \centering
  \hftwo\includegraphics[height=7.5cm]{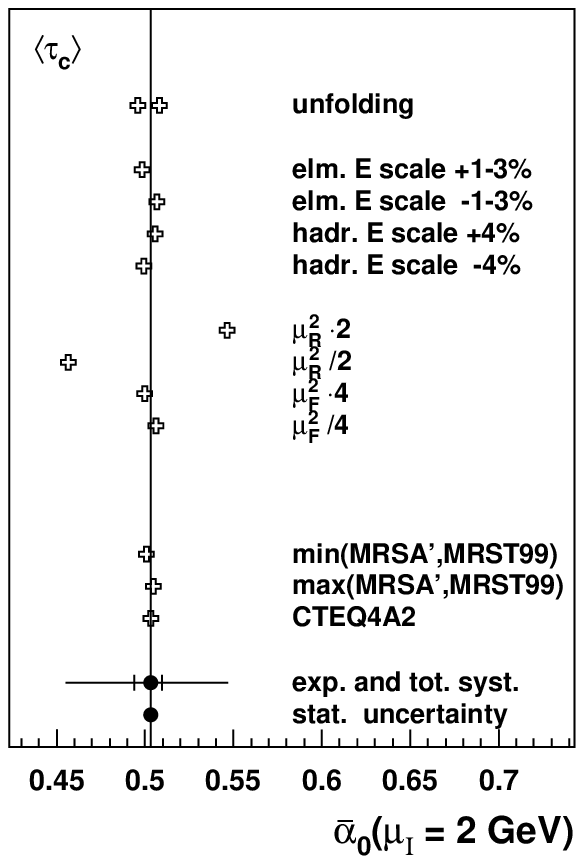}\hftwo%
  \includegraphics[height=7.5cm]{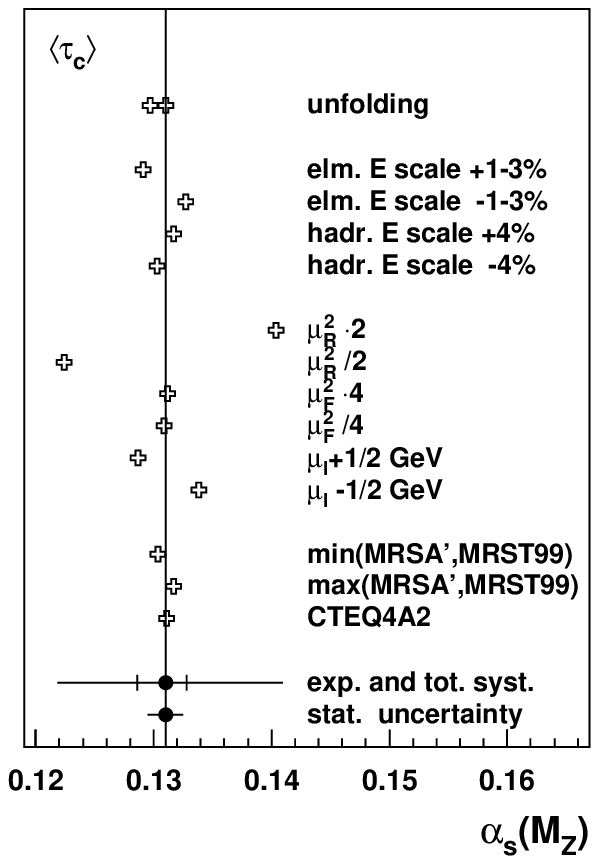}\hftwo\\
  \hftwo\includegraphics[height=7.5cm]{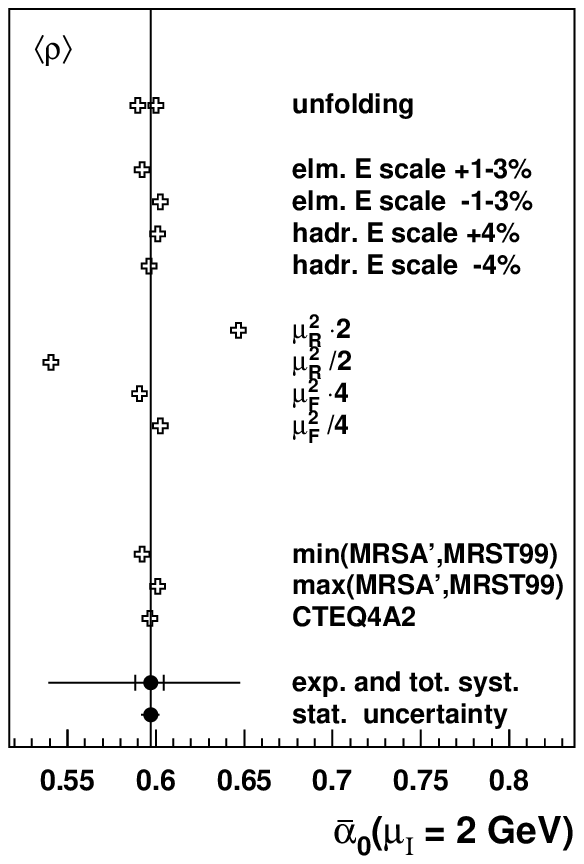}\hftwo%
  \includegraphics[height=7.5cm]{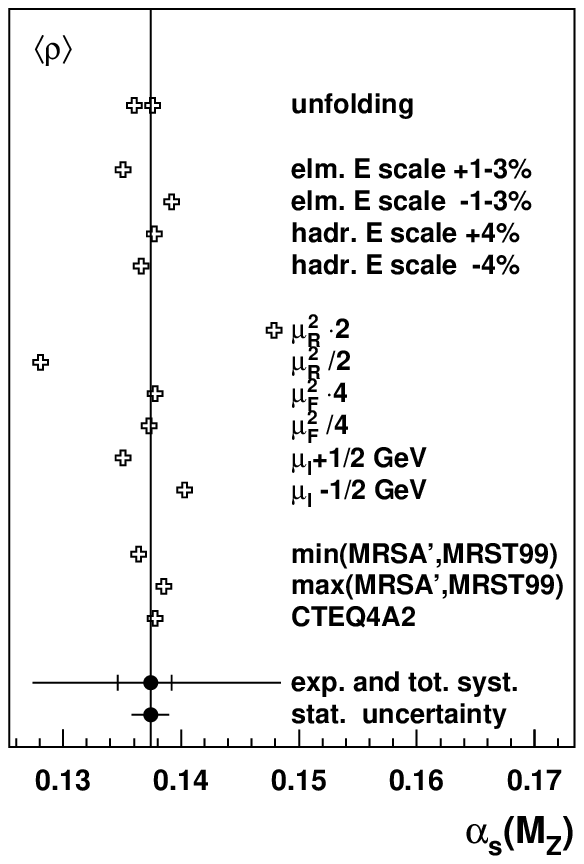}\hftwo\\
  \hftwo\includegraphics[height=7.5cm]{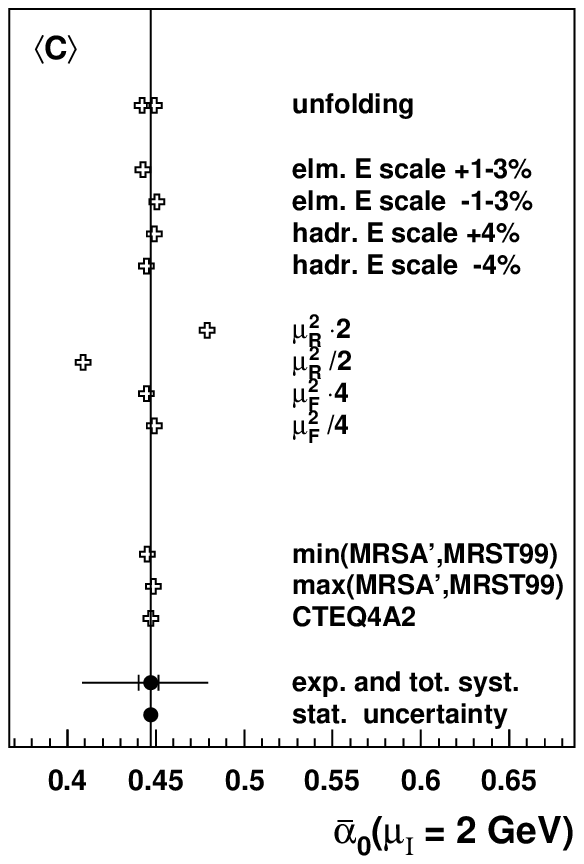}\hftwo%
  \includegraphics[height=7.5cm]{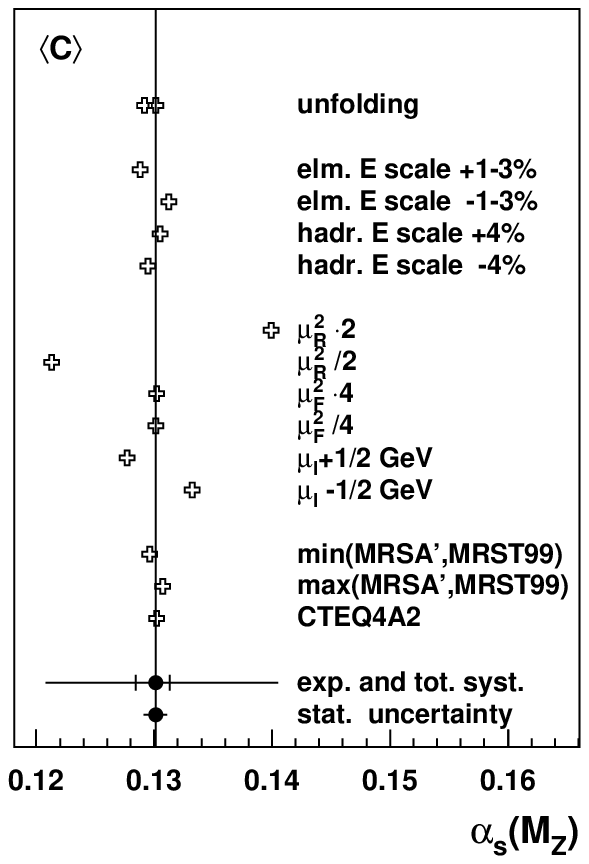}\hftwo
  \caption{Systematic uncertainties of $\anmi$ and $\asmz$ 
    for $\tau_C$ (top), $\rho$ (middle) and the $C$ parameter
    (bottom).}
  \label{fig:syserr2}
\end{figure}

The contributions of all systematic uncertainties are presented
graphically in figures~\ref{fig:syserr1} and~\ref{fig:syserr2} for
each of the five event shapes $\tau$, $B$, $\rho$, $\tau_C$ and $C$
where the coefficients $a_F$ are known.  Without such a prediction for
the two-jet event rates $y_{fJ}$ and $y_{k_t}$ the study of systematic
effects has been performed for experimental uncertainties only.


From figures~\ref{fig:syserr1} and~\ref{fig:syserr2} it is obvious
that $\tau$ and $B$ have different properties compared to $\rho$,
$\tau_C$ and $C$. For the latter the systematic uncertainties are
dominated by the variation of the renormalization scale.  In case of
$\tau$ and $B$ there is no prevailing source of systematics.  The
larger influence of experimental uncertainties can probably be related
to the explicit reference to the boson axis implied in the definitions
of $\tau$ and $B$.

Several additional cross-checks concerning systematic effects are
performed.  (i) Leptons pointing to partially inefficient regions in
the LAr calorimeter (see section~\ref{eventselection}) are removed.
This selectively diminishes the contribution to certain phase space
regions and is corrected for.  It is checked that the actual influence
is negligible even without unfolding.  (ii) The power correction
coefficients do not account for phase space constraints imposed on the
scattered lepton (cut no.\ 1).  Experimentally they are
unavoidable, but for testing purposes the pQCD calculations are repeated
without these cuts.  Except for $\mean{y_{k_t}}$ which increases by
$\sim 8\%$ at low $Q$ all other mean values change by less than $2\%$.
(iii) In the power corrections eqs.~(\ref{eqn:calp}),
(\ref{eqn:milan}) and~(\ref{eqn:bpow}) it is not obvious which
number of flavours to take.  As default $N_f = 5$ is used for the
perturbative part and the subtraction terms and $N_f = 3$ in the
coefficients of the power corrections.  Repeating the fits with $N_f =
3$ in both cases increases all $\an$ by $\approx 0.03$.  The values of
$\asmz$ are almost unaffected.

%%% Local Variables: 
%%% mode: latex
%%% TeX-master: "draft"
%%% End: 
      % QCD Analysis of Event Shape Means
\newpage
\section{Conclusions}

The event shape variables thrust, jet broadening, jet mass, $C$
parameter and two-jet event rates are studied in deep-inelastic $ep$
scattering.  The mean values exhibit a strong dependence on the scale
$Q$, which can be understood in terms of perturbative QCD and
non-perturbative hadronization effects decreasing with some power $p$
of $Q$.  The interest is to test whether the leading corrections to
perturbation theory can be parameterized without any assumptions on
the details of the hadronization process.  Simplistic models of the
form ${\rm const}/Q^p$ with $p=1$ or $2$ fail.  The concept of power
corrections in the approach initiated by Dokshitzer and Webber, which
predicts the power $p$ as well as the form and magnitude of the
hadronization contributions as a function of a new parameter $\ap$,
provides a much better and satisfactory description of the data.

The event shape variables defined in the current hemisphere of the
Breit frame --- $\tau$, $\tau_C$, $B$, $\rho$, and $C$ --- have
sizeable power corrections proportional to $1/Q$.  Two-parametric fits
yield for the non-perturbative parameter $\an \simeq 0.5$ within
$20\%$ supporting the notion of universal power corrections for the
event shapes.  The corresponding, correlated values of the strong
coupling $\asmz$ show a large spread incompatible within the
experimental uncertainties.  Possible explanations are missing higher
order QCD corrections, expected to be different for each event shape
variable, and/or incomplete knowledge of the power correction
coefficients.

Both two-jet rates $y$, defined for both hemispheres of the Breit system,
are almost
unaffected by hadronization effects.  For $y_{fJ}$ the conjectured
large positive contribution with $a_{fJ}=1$ is ruled out by the data,
which instead prefer small negative power corrections.  For $y_{k_t}$
fits with $p=1$ do not work properly.  The data are consistent with
the expectation of a power $p=2$, but unfortunately more quantitative
statements can not be made given the current experimental precision
and the unknown coefficient $a_{k_t}$.

The improved and extended analysis of mean event shape variables in
deep-inelastic scattering supports the concept of power corrections.
In order to achieve a better common description of the different event shapes
and to get a more coherent picture further theoretical progress is
needed.

Similar studies of power corrections, often combining data from
several experiments to cover a sufficiently large range in $Q$, have been
done for the analogous $e^+e^-$ event shape variables $\tau$, $B$,
$\rho$ and $C$~\cite{eedata}.  Remarkably, these analyses also find
universal parameters $\an \simeq 0.5$ within $20\%$.  But depending on
the choice of employed data sets correlated values of $\an$ and
$\asmz$ are found which deviate between the different determinations
by more than the quoted experimental accuracies. The situation appears
to be similar to the one in DIS.

{\em Note}\/: The experimental data of the event shape spectra shown
in figures~\ref{fig:dndFhl1} and~\ref{fig:dndFhl2} may be obtained in
tabular form from the HEPDATA data base at Durham.

%%% Local Variables: 
%%% mode: latex
%%% TeX-master: "draft"
%%% End: 
      % Conclusions
\newpage
\section*{Acknowledgements} 

We are grateful to the HERA machine group whose outstanding 
efforts have made and continue to make this experiment possible. We thank 
the engineers and technicians for their work in constructing and now 
maintaining the H1 detector, our funding agencies for financial support, the 
DESY technical staff for continual assistance, and the DESY directorate for 
the hospitality extended to the non--DESY members of the collaboration. 
We gratefully acknowledge valuable discussions with M.~Dasgupta,
Yu.L.~Dokshitzer and G.P.~Salam.

%%% Local Variables: 
%%% mode: latex
%%% TeX-master: "draft"
%%% End: 
              % Acknowledgements

%%% Local Variables: 
%%% mode: latex
%%% TeX-master: "draft"
%%% End: 
              % Bibliography

\end{document}